\documentclass[11pt]{article}
\usepackage{amsmath}
\usepackage{amsfonts}
\usepackage{amssymb}
\usepackage{amsthm}
\usepackage{epsfig}
\usepackage{graphicx}
\usepackage[ruled,vlined]{algorithm2e}
\usepackage{subfig}
\usepackage{graphicx}
\usepackage[countmax]{subfloat}
\usepackage{setspace}
\usepackage{subfig}
\usepackage{gensymb}
\usepackage{color}
\usepackage{bm}
\usepackage{authblk}
\usepackage{fullpage}
\usepackage{url}
\usepackage[authoryear]{natbib}
\usepackage{etoolbox}
\usepackage{amsthm}
\usepackage{multirow}
\newtheorem{lemma}{Lemma}

\theoremstyle{remark}

\makeatletter

\newif\ifabbreviation
\pretocmd{\thebibliography}{\abbreviationfalse}{}{}
\AtBeginDocument{\abbreviationtrue}

\begin{document}


\title{A Statistical Modeling Approach for \\ Spatio-Temporal Degradation Data}
\author[1]{Xiao Liu} 
\author[2]{Kyongmin Yeo} 
\author[2]{Jayant Kalagnanam}
\affil[1]{Department of Industrial Engineering\\ University of Arkansas}
\affil[2]{IBM Thomas J. Watson Research Center}

%
\date{ }

\maketitle

\vspace{0.5cm}
%
%



\begin{abstract}
This paper investigates the modeling of an important class of degradation data, which are collected from a spatial domain over time; for example, the surface quality degradation. Like many existing time-dependent stochastic degradation models, a special random field is constructed for modeling the spatio-temporal degradation process. In particular, we express the degradation at any spatial location and time as an additive superposition of two stochastic components: a dynamic spatial degradation generation process, and a spatio-temporal degradation propagation process. Some unique challenges are addressed, including the spatial heterogeneity of the degradation process, the spatial propagation of degradation to neighboring areas, the anisotropic and space-time non-separable covariance structure often associated with a complex spatio-temporal degradation process, and the computational issue related to parameter estimation. When the spatial dependence is ignored, we show that the proposed spatio-temporal degradation model incorporates some existing pure time-dependent degradation processes as its special cases. We also show the connection, under special conditions, between the proposed model and general physical degradation processes which are often defined by stochastic partial differential equations. A numerical example is presented to illustrate the modeling approach and model validation.

\end{abstract}


\section{Introduction} \label{sec:one}
Volume, Variety and Velocity are known as the three dimensions of Big Data. As new types of data rapidly emerge in today's industry, statistical techniques also need to evolve so as to cope with the additional complexity due to data diversity, which makes up the variety dimension of Big Data. In quality and reliability engineering, degradation data (e.g., the performance degradation of a device or system) provide a rich source of information for quality assurance and reliability assessment (Lu, Meeker and Escobar 1996; Meeker, Escobar and Lu 1998). In this paper, we investigate the modeling of a new type of degradation data: the spatio-temporal degradation data collected from a spatial domain over time.

One key observation arising from our industry practice is that, pure data-driven approaches are often limited even in data-rich engineering fields when operational insights are required based on engineering domain knowledge. In the age of Big Data, this issue will only become more relevant than ever when data-driven models are inevitably becoming more complex but less interpretable as both the volume and variety of data increase. Hence, in developing the spatio-temporal degradation model in this paper, we particularly focus on some important properties that a statistical model should have in the age of Big Data: 1) the computationally scalability for large data sets, and 2) the interpretability of the model based on underlying physics and engineering domain knowledge. We show the connection of the proposed statistical model to a large class of convection-diffusion processes that dominate a wide range of engineering applications. As to be discussed in the literature review, classical statistical spatio-temporal models hinge on the approximation of spatio-temporal processes by random fields with fully specified joint space-time covariance structures. But, for non-stationary and highly dynamic spatio-temporal processes, directly specifying space-time covariance structures is extremely ineffective, if not totally unrealistic at all. As a result, the specified covariance structure is usually motivated by mathematical tractability, and hence does not reflect the full complexity of the underlying physical processes. 

\subsection{Degradation Models based on Stochastic Processes} \label{sec:literature}

Degradation modeling for pure time-dependent degradation processes has been well studied over the past two decades (Meeker and Escobar 1998, Bae and Kvam 2004, Liu and Tang 2010, Ye and Chen 2014, Xu and Chen 2017), and the modeling of degradation data under dynamic environments has also received much attention in recent years (Liao and Tian 2012, Zhou, Serban and Gebraeel 2014, Bian et al. 2015, Hong et al., 2015).
In the literature, there exists an important class of models which describes degradation paths by a stochastic processes
\begin{equation}\label{eq:sp}
Y:[0,\infty )\rightarrow \mathbb{R},
\end{equation}
where $Y(t)$, a real-valued function, represents the value of degradation at time $t$. Such an idea can be traced back to the early work of Bhattaacharyya (1982) and Doksum and Hoyland (1992). Commonly used stochastic processes include those popular family members of L\'{e}vy processes, such as Wiener process (Tseng and Peng 2004), Gamma process (Singpurwalla 1995, Lawless and Crowder 2004) and Inverse Gaussian process (Ye and Chen 2014). The fundamental idea of using L\'{e}vy processes, with independent and stationary increments, is rooted in the assumption that every degradation increment can be seen as an additive superposition of a number of \emph{stationary and independent} small increments. Once the probability distribution for each small degradation increment is specified, the stochastic degradation process is uniquely determined. The use of stochastic processes allows us not only to model the temporal correlation structure of a degradation process, but also to leverage the well-established mathematical properties of those processes, such as the sample path properties and transition density.

Another advantage of stochastic degradation models is due to their natural connection to stochastic Partial Differential Equations (PDE), through which many physical models are defined. For example, the Paris-Erdogan law that describes the fatigue crack growth, the diffusion of lithium ions in solid and electrolyte phases that causes the output degradation of lithium-ion batteries, the gradual decomposition of polymer microspheres for pharmaceutical drug delivery over extended periods of time, etc. Since stochastic degradation models are based on the basic idea that every degradation increment is an additive superposition of a number of stationary and independent small increments, most of these models can be represented by a differential equation, $dY(t)  = udt + dB(t)$, with $dY(t)$, $u$ and $B(t)$ respectively being the degradation increment over an infinitesimal time interval $dt$, instantaneous degradation rate, and a stochastic process that fully characterizes the temporal correlation structure. One may refer to Nikulin et al. (2010) and Ye and Xie (2014) for a comprehensive review of stochastic degradation models under this framework.
%
%
\subsection{Degradation in Space and Time} \label{sec:degradationst}
In this paper, we investigate the modeling of a new type of degradation data which are collected not only over time but also from a spatial domain. Following the main idea behind stochastic degradation models discussed above, our goal is to extend the existing time-dependent degradation process, $Y:[0,\infty )\rightarrow \mathbb{R}$, to a space-time process,
\begin{equation}\label{eq:stp}
Y:\mathbb{R}^d \times [0,\infty )  \rightarrow \mathbb{R},
\end{equation}
where $Y$ is now a spatio-temporal random field, and $d \in \mathbb{N}^+$ is the dimension of the spatial domain.

Degradation data of this kind usually exhibit complex correlation structure in space and time, making it no longer appropriate to model the data collected from different spatial locations independently using existing methods. Some examples of spatio-temporal degradation include the aging of asphalt pavement on highways, the corrosion of oil and gas pipelines, the bulk erosion when water diffuses into a polymer structure, the surface quality degradation/drifting of a manufacturing process such as the Wafer Intelligence Scanner (WIS) data that measure the RGB color of a silicon wafer layer. In Appendix C, we also show a real spatio-temporal model that describes the degradation of polymer microspheres composed of poly(D,L-lactic-co-glycolic acid) (Versypt et al. 2015).

\begin{figure}[!htb]
	\begin{center}
		\includegraphics[width=1\textwidth]{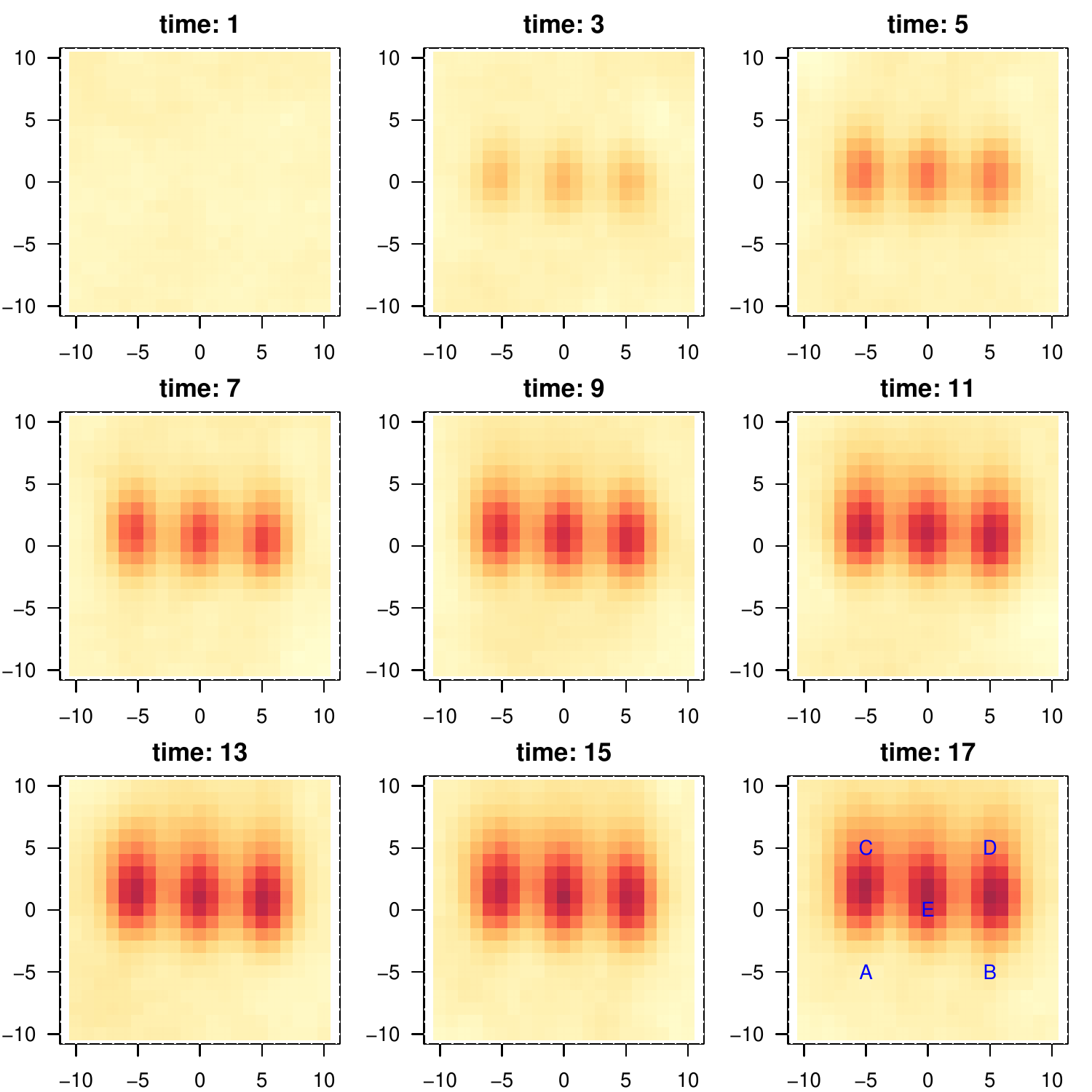}
		\caption{Degradation on a two-dimensional surface at equally spaced time intervals. Three regions, aligned horizontally, are clearly observed with higher degradation, and the degradation propagation along the south-to-north direction can also be seen.} \label{fig:example1}
	\end{center}
\end{figure}

Figure \ref{fig:example1}, as a motivating example, shows the degradation measured on a two-dimensional surface over 9 equally spaced time intervals (the darker the color, the higher the degradation). Due to the confidentiality agreement, the actual application is not mentioned throughout this paper. In this example, the degradation data are measured by a surface scanner and aggregated to a $21 \times 21$ pixel array, and observations are available from time 1 to time 20. At time 1 (the initial condition), the measured degradation over the entire surface is relatively low, and the spatial variation across different pixels appears to be random with no special spatial pattern. Starting from time 3 and time 5, three regions, aligned horizontally in the center of the surface, are observed with higher degradation. Both the degradation level and the size of these three regions gradually grow over time. 

Compared to the modeling of pure time-dependent degradation data, the modeling of spatio-temporal degradation data of this kind poses some common challenges. Firstly, the degradation may propagate within the spatial domain along certain directions (e.g., south-to-north direction in Figure \ref{fig:example1}). This special phenomenon, which is common for many engineering applications such as surface erosion, needs to be addressed by the spatio-temporal degradation model. Because the propagation of degradation often presents a dominant direction, the measured degradation data over a spatial domain are directionally dependent, as opposed to isotropy which implies identical properties in all directions. From the statistical modeling point of view, if a random field is to be constructed to model the degradation data, the spatio-temporal correlation structure of that field needs to be anisotropic, making the modeling much more challenging.

Secondly, over a spatial domain, the degradation at different locations often exhibits a certain level of heterogeneity. The heterogeneity is due to not only the anisotropic degradation propagation discussed above, but also the difference in degradation rate at different spatial locations, governed by some underlying mechanism. In Figure \ref{fig:example1}, we observe that the three high-degradation regions have significantly higher degradation rate than other areas within the domain. The degradation paths at different locations, say, the five locations marked by uppercase letters, ``A'', ``B'', ``C'', ``D'' and ``E'', in the bottom-right subplot of Figure \ref{fig:example1}, must have very different degradation progression over time.


Thirdly, the spatial propagation, anisotropy, and heterogeneity eventually lead to an extremely complex spatio-temporal correlation structure. For pure time-dependent stochastic degradation models, the temporal correlation structure can be directly specified, such as Brownian, Gamma and Inverse-Gaussian, and easily validated using data. For the modeling of spatio-temporal degradation data, however, one needs to choose not only the temporal correlation but also the spatial correlation structure. Furthermore, the correlation is often anisotropic and space-time non-separable, meaning that the spatio-temporal covariance function cannot be expressed as the multiplication of a spatial covariance function and a temporal covariance function. There have been prolonged interests in the spatio-temporal statistics to provide flexible and effective ways to construct covariance functions (Cressie and Huang 1999, Gneiting 2002, Banerjee et al., 2004, Fuentes et al., 2005, Reich et al., 2011, Liu et al., 2018). For many real engineering applications, the space-time covariance structures can be extremely complicated due to the interactions between the spatial variation of degradation rate and degradation propagation over the spatial domain. It is challenging to specify appropriate space-time covariance functions that adequately model such complicated covariance structures (Calder 2007, Ghosh et al., 2010).

The last challenge worth mentioning is the computational cost associated with parameter estimation. For large spatio-temporal degradation data sets, estimating the unknown parameters of a spatio-temporal process can be computationally costly, if not impossible at all. For example, the Maximum Likelihood Estimation (MLE) typically requires an $\mathcal{O}((N_t \times N_s)^3)$ operation for the inversion of a large covariance matrix, where $N_t$ and $N_s$ are the total number of sampling times and locations, known as the ``big n problem'' in spatial statistics. In addition, the total number of parameters in a spatio-temporal degradation model is much larger than that of a pure time-dependent degradation model. Hence, maintaining the computational tractability for the proposed spatio-temporal degradation model is critical in order to make the proposed method useful in practice.

\subsection{Overview of the Paper} \label{sec:overview}
Section \ref{sec:two} presents the statistical model that describes the spatio-temporal degradation process by a random field with a special space-time covariance structure.
Section \ref{sec:link1} establishes the link, under special conditions, between the proposed spatio-temporal degradation model and existing time-dependent degradation models, as well as the link between the proposed statistical model and PDE-based physical models.
Parameter estimation is presented in Section \ref{sec:MLE} where the motivating example in Section \ref{sec:degradationst} is re-visited to demonstrate the proposed modeling approach. In addition, we provide discussions on how the model validation and selection can be done based on data, as well as how the distributions of the First-Passage-Time and the First-Passage-Location can be approximated using simulation.
Section \ref{sec:conclusion} concludes the paper and highlights some future research directions.

\section{A Spatio-Temporal Degradation Model} \label{sec:two}
\subsection{The Basic Framework} \label{sec:model1}
We first consider a discrete-in-time and continuous-in-space spatio-temporal random field, $\{Y(\bm{s},t); \bm{s} \in \mathbb{R}^d, t \in \mathbb{N}^+\}$, where $Y(\bm{s},t)$ represents the degradation at time $t$ and location $\bm{s}$ in a \emph{d}-dimensional space. Without loss of generality, we let $d=2$ and the results presented in this paper can be extended to higher dimensions. As discussed in Section \ref{sec:literature}, the fundamental idea behind existing stochastic degradation models is rooted in the assumption of additive accumulation of degradation. Following the same idea, we assume that $Y(\bm{s},t)$ takes an additive form as follows:
\begin{equation}\label{eq:Y}
Y(\bm{s},t) = G_{\Delta}(\bm{s},t) + Z(\bm{s},t),
\end{equation}
where
\begin{equation}\label{eq:G}
G_{\Delta}(\bm{s},t) = g_{\Delta}(\bm{s},t) + \varepsilon_{\Delta}(\bm{s},t)
\end{equation}
with $Z(\bm{s},t)$ being a stochastic process depending on $Y(\bm{s},t-\Delta)$.

The first term $G_{\Delta}(\bm{s},t)$ on the right hand side of (\ref{eq:Y}) is a spatial process that represents the amount of degradation \emph{generated} at location $\bm{s}$ over the time interval $(t-\Delta,t]$ with length $\Delta$. Apparently, the modeling of spatio-temporal degradation data requires us to consider the spatial variation of degradation rate, while such a problem is irrelevant to the modeling of time-dependent degradation data.
Consider, for example, the surface degradation where degradation is only initiated from some locations. To capture the spatial correlation of the generation of degradation within the spatial domain and account for the uncertainty over the time interval $(t-\Delta,t]$, $G_{\Delta}(\bm{s},t)$ is decomposed in (\ref{eq:G}) as the sum of a deterministic mean-value term $g_{\Delta}(\bm{s},t)$ with (possibly time-varying) covariates, and a spatial process $\varepsilon_{\Delta}$ with covariance function $c_{\Delta}(\cdot)=\Delta \cdot c(\cdot)$, where $c(\cdot)$ is the covariance function of a white-in-time Gaussian random field. We leave the detailed modeling of $g_{\Delta}(\bm{s},t)$ to Section \ref{sec:linear}.

The second random term $Z(\bm{s},t)$, which depends on $Y(\bm{s},t-\Delta)$, captures \emph{the propagation of degradation over space and time}.
Some typical examples include the surface corrosion, crack propagation, etc., where degradation is initiated at some locations and then propagated to neighboring areas along certain directions. It is possible to see that, the stochastic propagation process, $Z(\bm{s},t)$, to a large extent determines the spatio-temporal correlation structure of the random process $Y(\bm{s},t)$.
Brown et al.(2000) advocated the use of convolution to approximate the propagation phenomenon in space under a constant vector field. Recently, Sigrist et al.(2015) presented a spatio-temporal Gaussian process directly derived from the solution of a stochastic PDE describing the convection-diffusion process, and proposed to use a Fourier spectral method for efficient computation. In the same spirit, Calder (2007) proposed a dynamic space-time model in which the quantity of interest is expressed as the convolution of a latent process. Such a model fits into the general dynamic modeling framework for spatial-temporal data introduced in Stroud et al. (2001).

Following the work of Wikle and Cressie (1999), Brown et al.(2000) and Sigrist et al.(2015), we adopt the idea that the amount of degradation propagated to a certain location at time $t$ can be expressed as a linear combination of the degradation at neighboring locations at time $t-\Delta$, weighted by some spatial kernel function. Hence, a convolution model, with a Gaussian kernel, is used to describe the propagation process as follows,
\begin{eqnarray}
\begin{split}
Z(\bm{s},t)  & = \zeta_\Delta \left\{\omega_{\Delta} * Y(\bm{s},t-\Delta) \right\} \\ & = \zeta_\Delta \left\{ \int_{\mathbb{R}^2} \omega_{\Delta}(\bm{x}) Y(\bm{s}-\bm{x},t-\Delta)d \bm{x} \right\} \\ & = \zeta_\Delta \left\{ \omega_{\Delta} * Z(\bm{s},t-\Delta) + \omega_\Delta * g_{\Delta}(\bm{s},t-\Delta)  + \omega_{\Delta} * \varepsilon_{\Delta}(\bm{s},t-\Delta) \right\},
\label{eq:Z}
\end{split}
\end{eqnarray}
where $*$ denotes the convolution operation, $\omega_\Delta$ is the convolution kernel which is an infinitely divisible function, and $\zeta_\Delta<1$ is a scaling factor. In particular, we respectively define the scaling factor $\zeta_\Delta$ and the convolution kernel $\omega_\Delta$  as
\begin{align}
\zeta_\Delta &= \exp(-\lambda \Delta), \quad \lambda>0, \label{eq:Z1} \\
\omega_\Delta(\bm{x}) & = \frac{1}{2\pi|\bm{\Sigma}_{\Delta}|^{1/2}}\exp\left\{-\frac{(\bm{x}-\bm{\mu}_{\Delta})^{\top} \bm{\Sigma}_{\Delta}^{-1}(\bm{x}-\bm{\mu}_{\Delta})}{2}\right\} \nonumber \\
& = \phi(\bm{x}; \bm{\mu}_{\Delta}, \bm{\Sigma}_{\Delta}), \label{eq:Z2}
\end{align}
where $\phi$ denotes the probability density function of a bivariate Gaussian distribution, $\bm{\mu}_{\Delta}$ reflects the direction and speed of the spatial propagation of degradation over a time interval, and $\bm{\Sigma}_\Delta$ is the covariance matrix of the Gaussian convolution kernel. 
The choice of the convolution model (\ref{eq:Z}) as well as (\ref{eq:Z1}) and (\ref{eq:Z2}) is validated using data in Section \ref{sec:numerical}, and is also justified in Section \ref{sec:link1} by establishing the link between the proposed spatio-temporal model and existing time-dependent degradation models, as well as the link between the proposed statistical model and physical degradation models given by stochastic PDE.


The convolution model implies that the amount of degradation propagated to a location $\bm{s}$ over a time interval is a linear combination of the degradation in the neighborhood of $\bm{s}$, given the direction and speed of the spatial propagation of degradation. Let $\bm{v}=(v_1,v_2)$ be a vector that specifies the direction and speed of the spatial propagation of degradation, we assume the following parameterization:
\begin{eqnarray}
\bm{\mu}_\Delta = \bm{v}\Delta,
\label{eq:mu}
\end{eqnarray}
which is the propagation distance along the direction of $\bm{v}$ over a time interval $\Delta$, and
\begin{eqnarray}
\bm{\Sigma}_{\Delta} = \bm{R}^{-1}
\begin{pmatrix}
\rho_1 \Delta & 0\\
0 & \rho_2 \Delta
\end{pmatrix} (\bm{R}^{\top})^{-1},
\label{eq:sigma}
\end{eqnarray}
where $\bm{R}$ is proper rotation matrix (i.e., $\bm{R}^{\top}=\bm{R}^{-1}$ and $\mathrm{det}\bm{R}=1$) given by:
\begin{eqnarray}
\bm{R} =
\begin{pmatrix}
\cos(\alpha_{\bm{v}}) & -\sin(\alpha_{\bm{v}})\\
\sin(\alpha_{\bm{v}}) & \cos(\alpha_{\bm{v}})
\end{pmatrix}
\label{eq:R}
\end{eqnarray}
with $\alpha_{\bm{v}}\in[0,2\pi)$ being the counter-clockwise rotation angle of the propagation vector $\bm{v}$ from the horizontal axis. Hence, the parameters, $\rho_1$ and $\rho_2$, respectively control the standard deviations of the convolution kernel $\omega_\Delta$ in directions which are parallel and perpendicular to the direction of degradation propagation.

Further discretizing $\Delta$ in (\ref{eq:Z}) into a number of $n$ small time intervals ($\delta = \Delta/n$) gives
\begin{eqnarray} \label{eqn:discretized}
\begin{split}
Z(\bm{s},t) = & \sum_{i=1}^{n}\exp(-i \lambda \delta) \omega_{\delta}^{*i} * g_{\delta}(\bm{s},t-i\delta) \\
&+ \exp(-n \lambda \delta) \omega_{\delta}^{*n} * Z(\bm{s},t-\Delta)
+ \sum_{i=1}^{n}\exp(-i \lambda \delta) \omega_{\delta}^{*i} * \varepsilon_{\delta}(\bm{s},t-i\delta)
\label{eq:Z3}
\end{split}
\end{eqnarray}
with $*n$ denoting the $n$-fold convolution operation. Substituting (\ref{eq:Z3}) into (\ref{eq:Y}) and noting that the convolution of Gaussians is still a Gaussian, we obtain the expression of the degradation over continuous space and discrete time:
\begin{eqnarray}
\begin{split}
Y(\bm{s},t)  & =  \sum_{i=0}^{n} \left\{ \exp(-i\lambda\delta) \phi_{i} \left( \bm{s} \right) * g_{\delta}(\bm{s},t-i\delta) \right\} \\ &  + \sum_{i=0}^{n} \left\{ \exp(-i\lambda\delta) \phi_{i} \left( \bm{s} \right) * \varepsilon_{\delta}(\bm{s},t-i\delta)\right\}  \\ & + \exp(-n\lambda\delta) \phi_{n} \left( \bm{s}\right) * Z(\bm{s},t-n\delta) .
\label{eq:Y1_0}
\end{split}
\end{eqnarray}
where
\begin{eqnarray}
\phi_{i} \left( \bm{s} \right) = \begin{cases} \phi \left( \bm{s}; i\bm{\mu}_{\delta}, i \bm{\Sigma}_{\delta} \right),& \quad i \in \mathbb{N}^+ \\ \kappa\left( \bm{s} \right), & \quad i=0  \end{cases}
\end{eqnarray}
with $\kappa$ being a Dirac delta function. The first term on the right hand side of (\ref{eq:Y1_0}) captures the deterministic amount of degradation generated over the interval $(t-\Delta, t]$, the second term captures the uncertainty associated with the degradation generated over the interval $(t-\Delta, t]$, while the last term captures the effects of degradation accumulated prior to time $t-\Delta$.

When $\lambda>0$ and $n \rightarrow \infty$, $Y(\bm{s},t)$ is approximated by a stationary spatio-temporal random field as follows:
\begin{eqnarray}
\begin{split}
Y(\bm{s},t)  & \approx \sum_{i=0}^{\infty} \left\{ \exp(-i\lambda\delta) \phi_{i} \left( \bm{s} \right) * g_{\delta}(\bm{s},t-i\delta) \right\} \\ &  + \sum_{i=0}^{\infty} \left\{ \exp(-i\lambda\delta) \phi_{i} \left( \bm{s} \right) * \varepsilon_{\delta}(\bm{s},t-i\delta)\right\}.
\label{eq:Y1}
\end{split}
\end{eqnarray}

The approximation error is given by the difference between (\ref{eq:Y1}) and (\ref{eq:Y1_0}), which is stochastically bounded as $\mathcal{O}_p(e^{-n\lambda\delta})$. Interesting readers may refer to Appendix A.

Let
\begin{eqnarray}
\Psi_{i}(\bm{s}) = \exp(-i\lambda\delta) \phi_{i} \left( \bm{s} \right),
\label{eq:Psi}
\end{eqnarray}
the following result fully characterizes the spatio-temporal correlation of the process defined in (\ref{eq:Y1}).

\begin{lemma}
	For the stochastic degradation process defined in (\ref{eq:Y1}), the covariance of the degradation between $(\bm{s}_1,t_1)$ and $(\bm{s}_2,t_2)$ (assuming $t_2-t_1=\jmath \delta$ for some $\jmath=0,1,2,...$) is given by
	\begin{eqnarray}
	\begin{split}
	\mathrm{cov}(Y(\bm{s}_1,t_1),Y(\bm{s}_2,t_2)) & = \sum_{i=0}^{\infty} ( \tilde{\Psi}_{i} * \Psi_{\jmath+i,t_2} * c_{\delta} )(\bm{d}) \\ & + I_{\{\jmath=0\}}c_{\delta}(\bm{d})
	\label{eq:cov}
	\end{split}
	\end{eqnarray}
	where $\bm{d}=\bm{s}_2-\bm{s}_1$, $\tilde{\Psi}_{i}(\bm{s}) \equiv \Psi_{i}(-\bm{s})$, and $I_{\{\jmath=0\}}=1$ only when ${\jmath=0}$, otherwise, $I_{\{\jmath=0\}}=0$.
\end{lemma}

The derivation of (\ref{eq:cov}) is provided in Appendix B. Note that, since the random field $\varepsilon(\bm{s},t)$ is isotropic, $c(\bm{d})=c(||\bm{d}||)$ with $||\bm{d}||$ being the distance between $\bm{s}_1$ and $\bm{s}_2$. It is seen from (\ref{eq:cov}) that the covariance, $\mathrm{cov}(Y(\bm{s}_1,t_1),Y(\bm{s}_2,t_2))$, is determined by not only the separation of time and space, but also the degradation propagation speed and direction between times $t_1$ and $t_2$. Note that, one key assumption of the proposed model is that the propagation field is uniform in space and invariant in time. Only under this assumption can we approximate the stochastic part of the degradation process by a stationary process, and obtain the covariance structure shown in Lemma 1.

\subsection{A Linear Representation} \label{sec:linear}
We show in this section that, the spatio-temporal degradation process $\{ Y(\bm{s},t) \}$ in (\ref{eq:Y1}) has a linear representation, if the deterministic degradation generation at location $\bm{s}$ over the time interval $(t-\Delta,t]$, $g_{\Delta}(\bm{s},t)$, is a linear function of its (possibly time-varying) covariates.

Let
\begin{eqnarray}
g_{\Delta}(\bm{s},t)= \bm{x}_0(\bm{s},t) \bm{\beta}^{\top},
\label{eq:glinear}
\end{eqnarray}
where $\bm{x}_0(\bm{s},t)=(x^{(1)}(\bm{s},t),\ldots,x^{(k)}(\bm{s},t))$ is a row vector of length $k$ that contains covariates, and $\bm{\beta}=(b^{(1)},\ldots,b^{(k)})$ a row vector of parameters that determines the effects of covariates on the degradation generation over the time interval $(t-\Delta,t]$. Note that, the proposed model allows $\bm{x}_0(\bm{s},t)$ to vary over time, and this is often the case when the covariates depend on dynamic environmental conditions.

Substituting the expression of $g_{\Delta}(\bm{s},t)$ into (\ref{eq:Y1}), the expected degradation at location $\bm{s}$ and time $t$ is obtained:
\begin{eqnarray}
\begin{split}
\mathbb{E}(Y(\bm{s},t)) & \approx  g(\bm{s},t) + \sum_{i=1}^{n} \left\{ \Psi_{i}(\bm{s}) * (\bm{x}_0(\bm{s},t-i\delta)\bm{\beta}^{\top})\right\}\\
& = \bm{x}_0(s,t) \bm{\beta}^{\top}  + \sum_{i=1}^{n} \left\{ \sum_{p=1}^{k} b^{(p)} \left[ \Psi_{i}(\bm{s}) * x^{(p)}(\bm{s},t-i\delta) \right]  \right\}.
\label{eq:EY}
\end{split}
\end{eqnarray}

Let $\bm{x}_{i}(\bm{s},t)=(\Psi_{i}(\bm{s})*x^{(1)}(\bm{s},t-i\delta),\ldots,\Psi_{i}(\bm{s})*x^{(k)}(\bm{s},t-i\delta))$ for $i=0,\ldots,n$, and recall that $\Psi_{i}(\bm{s})=1$ if $i=0$, (\ref{eq:EY}) is further simplified to a linear form:
\begin{eqnarray}
\mathbb{E}(Y(\bm{s},t)) = \left( \sum_{i=0}^{n}\bm{x}_{i}(\bm{s},t) \right) \bm{\beta}^{\top} = \tilde{\bm{x}}(\bm{s},t)\bm{\beta}^{\top}.
\label{eq:EY2}
\end{eqnarray}
with $\left( \sum_{i=0}^{n}\bm{x}_{i}(\bm{s},t) \right)=\tilde{\bm{x}}(\bm{s},t)$.

Equation (\ref{eq:EY2}) can be interpreted as follows: the covariates of $g_{\Delta}(\bm{s},t)$ are firstly transformed by the convolution operation, and the expected degradation is given by a linear combination of the transformed covariates. Since convolution is a linear operation, the linear form of (\ref{eq:EY2}) is expected when $g_{\Delta}(\bm{s},t)$ is a linear function of its covariates. Finally, suppose that degradation is measured at $N_s$ locations and $N_t$ sampling times, we obtain from (\ref{eq:EY2}) a linear form as follows:
\begin{eqnarray}
\bm{Y} = \bm{X} \bm{\beta} + \bm{e},
\label{eq:YLinear}
\end{eqnarray}
where $\bm{Y}=(Y(\bm{s}_1,t_1),\ldots,Y(\bm{s}_{N_s},t_1),\ldots,Y(\bm{s}_{N_s},t_{N_t}))^{\top}$
is a column vector of length $N_s \times N_t$ and
$\bm{X}=(\tilde{\bm{x}}(\bm{s}_1,t_1),\tilde{\bm{x}}(\bm{s}_2,t_1),\ldots,\tilde{\bm{x}}(\bm{s}_{N_s},t_1),\ldots,\bm{x}(\bm{s}_{N_s},t_{N_t}))^{\top}$
is a $N_sN_t \times k$ matrix, and $\bm{e}$ is the error process with mean zero and variance $\bm{\Sigma}_{\bm{Y}}$ given by (\ref{eq:cov}).

\subsection{The Model under Continuous Space and Continuous Time Domain} \label{sec:model2}

To obtain the stochastic degradation process under continuous space and continuous time, $\{Y(\bm{s},t); \bm{s} \in \mathbb{R}^d, t \in \mathbb{R}\}$, we extend the stochastic degradation process defined in (\ref{eq:Y1}). First, it is necessary to assume that there exists a real-value function $u(\bm{s},t)$ that satisfies:
\begin{eqnarray}
g_{\delta}(\bm{s},t)  = \int_{t-\delta}^{\delta} u(\bm{s},x) dx.
\label{eq:u}
\end{eqnarray}
Since $g_{\delta}(\bm{s},t)$ is previously defined as the total amount of degradation generated at location $\bm{s}$ and over the time interval $(t-\delta,t]$. Hence, $u(\bm{s},t)$ can be naturally interpreted as the degradation generation rate at location $\bm{s}$ and time $t$.

Similarly, we also assume that there exists a spatial random process such that the following stochastic integral holds:
\begin{eqnarray}
\varepsilon_{\delta}(\bm{s},t)  = \int_{t-\delta}^{\delta} d\tau(\bm{s},x) dx.
\label{eq:B}
\end{eqnarray}
As shown in Brown et al.(2000) , if $\delta^{-1}\varepsilon_{\delta}(\cdot,t)$ is asymptotically Gaussian with mean zero and covariance $c_{\tau}(\cdot)$, it is easy to see that $\tau(\bm{s},t)$ in (\ref{eq:B}) is a spatially correlated Brownian motion such that $d\tau(\cdot,t) \sim N(0, c_{\tau}(\cdot)dt)$.

Hence, by letting $n \rightarrow \infty$ and $\delta \rightarrow 0$, we obtain from (\ref{eq:Y1}) the expression of the stochastic degradation process under continuous space and time:
\begin{eqnarray}
\begin{split}
Y(\bm{s},t)  & = (\Psi^{(c)} *_{\bm{s}} *_t u)(\bm{s},t) + (\Psi^{(c)} *_{\bm{s}} *_t d\tau)(\bm{s},t) \\ & = (\Psi^{(c)} *_{\bm{s}} *_t (u+d\tau))(\bm{s},t)
\label{eq:YContinuous}
\end{split}
\end{eqnarray}
where $*_{\bm{s}}$ and $*_t$ are respectively the convolution with respect to space $\bm{s}$ and time $t$, and $\Psi^{(c)}$, defined as follows, is the continuous version of $\Psi_{i}(\bm{s})$ in (\ref{eq:Psi}):
\begin{eqnarray}
\Psi^{(c)}(\bm{s},t) = \begin{cases} \exp(-\lambda t) \phi \left( \bm{s}; \bm{\mu}_t,\bm{\Sigma}_t)  \right),& \quad t \geq 0 \\
0, & \quad t<0  \end{cases}
\label{eq:PsiContinuous}
\end{eqnarray}

It is important to note that, $u+d\tau$ in (\ref{eq:YContinuous}) can be interpreted as the stochastic degradation rate which is expressed as the sum of a deterministic degradation rate $u$ and a spatial error term $d\tau$. In fact, $u+d\tau$ is just the continuous version of equation (\ref{eq:G}), and equation (\ref{eq:YContinuous}) is the continuous version of (\ref{eq:Y}) and (\ref{eq:G}). More importantly, (\ref{eq:YContinuous}) implies that the degradation at location $\bm{s}$ and time $t$ linearly depends, through the convolution operations in both space and time, on the degradation generated in the entire spatial domain and over the entire history $[0,t]$.

\section{The Link to Time-Dependent and Physical Degradation Models} \label{sec:link1}
In this section, we show the connection between the proposed spatio-temporal degradation model and existing pure time-dependent stochastic degradation models, as well as the connection between the proposed statistical model and physical degradation models.
\subsection{The Connection to Time-Dependent Degradation Models}
As discussed in the introduction section, existing stochastic degradation models can often be expressed by differential equations as follows:
\begin{eqnarray}
dY(t)  = u dt + dB(t),
\label{eq:degradation}
\end{eqnarray}
where $Y(t)$ is the degradation over time, $u$ is the degradation rate, and $B(t)$ is some stochastic process; see Ye and Xie (2014) for a comprehensive review on stochastic degradation models under this framework. For example, if $B(t)$ is a Brownian process, (\ref{eq:degradation}) becomes the widely used Wiener degradation model.

Following the approach described in Brown et al.(2000) , we can also obtain the SPDEs representation of the spatio-temporal degradation model (\ref{eq:YContinuous}):
\begin{eqnarray}
dY(\bm{s},t)  = ( u(\bm{s},t) - \frac{1}{2}\{ \Gamma Y(\cdot,t) \}(\bm{s}) )dt + d\tau(\bm{s},t),
\label{eq:SPDE}
\end{eqnarray}
where $\Gamma$ is a spatial linear operator defined as:
\begin{eqnarray}
\{\Gamma f(\cdot)\}(\bm{s}) = \frac{\partial}{\partial \bm{s}^{T}}f(\bm{s})\bm{\mu} - \mathrm{trace}\left\{ \frac{\partial^2}{\partial \bm{s} \partial \bm{s}^{T}}f(\bm{s}) \right\}\bm{\Sigma} + 2\lambda f(\bm{s}),
\label{eq:Gamma}
\end{eqnarray}
and $\tau(\bm{s},t)$, previously defined in (\ref{eq:B}), is a spatially correlated Brownian motion such that $d\tau(\cdot,t) \sim N(0, c_{\tau}(\cdot)dt)$. Here, $c_{\tau}(\cdot) = \lim_{\delta\rightarrow 0} (\delta^{-1}c_{\delta}(\cdot))$.

It is possible to see that,  the first term $u(\bm{s},t)$ in (\ref{eq:SPDE}) describes the degradation rate at location $\bm{s}$ and at time $t$, which corresponds to the term $u$ in (\ref{eq:degradation}). The second term $\frac{1}{2}\{ \Gamma Y(\cdot,t) \}(\bm{s})$ captures the propagation of degradation over space, which of course does not appear in the traditional degradation model (\ref{eq:degradation}). The last stochastic term $d\tau(\bm{s},t)$ in (\ref{eq:SPDE}) describes the uncertainty associated with the degradation process, which naturally connects to $dB(t)$ in (\ref{eq:degradation}). Next, we consider two special scenarios:

\begin{itemize}
	\item \textbf{The process without spatial degradation propagation and decay}. If the spatial degradation propagation and decay are ignored (i.e., $\bm{\mu}=\bm{\Sigma}=\lambda=0$), the second term of (\ref{eq:SPDE}) vanishes, and the degradation process (\ref{eq:SPDE}) becomes
	\begin{eqnarray}
	dY(\bm{s},t)  = u(\bm{s},t)dt + d\tau(\bm{s},t),
	\label{eq:SPDEcase1}
	\end{eqnarray}
	which is exactly a classic multivariate Wiener degradation process. For such a process, the correlation structure of this process is fully determined by the choice of $c(\bm{s})$. In the context of spatial statistics, any general spatial covariance functions can be used such as Exponential, Mat\'{e}rn and Gaussian (Cressie and Huang 1999). In particular, since $\tau(\bm{s},t)$ is a spatially correlated Brownian motion (i.e., a multivariate Brownian motion), it is immediately seen that the (marginal) degradation process at any location $\bm{s}$ becomes a traditional Brownian degradation process.
	\item \textbf{The process without spatial degradation propagation}. If only the spatial degradation propagation is ignored (i.e., $\bm{\mu}=\bm{\Sigma}=0$ and $\lambda > 0$), the degradation process (\ref{eq:SPDE}) becomes
	\begin{eqnarray}
	dY(\bm{s},t)  = ( u(\bm{s},t) - \lambda Y(\bm{s},t) )dt + d\tau(\bm{s},t).
	\label{eq:SPDEcase2}
	\end{eqnarray}
	From Sigrist et al.(2015), the process above is a space-time separable random process with covariance function $(2\lambda)^{-1}\exp(-\lambda |t|)c(\bm{s})$.
\end{itemize}

\subsection{The Connection to Physical Models}
For real engineering problems, physical degradation models are typically defined in the form of Partial Differential Equations (PDE); for example, the reaction-diffusion model for autocatalytic degradation in polymer microspheres (Versypt et al. 2015), the image quality degradation (Foyer and Zou 2006), etc. Hence, it is of a great practical importance to establish the connection between the proposed statistical model and physical models given by PDE. Such a connection not only helps to justify the proposed statistical approach, such as the choice of the convolution approach with a Gaussian kernel (\ref{eq:Z}), but also demonstrate how a physics-based statistical degradation model can be constructed for real problems; also see Doganaksoy and Hall (2013).

In order to maintain the clarity and readability of the paper, the main results in this section are presented in Appendix C. In particular, we first establish the connection between the proposed model and a general PDE with convection, diffusion, decay and generation terms. Then, we illustrate such a connection using a real reaction-diffusion model that describes the gradual degradation (decomposition) of polymer microspheres composed of poly(D,L-lactic-co-glycolic acid) (PLGA). 

\section{Parameter Estimation} \label{sec:MLE}
The proposed spatial-temporal degradation model contains a much larger number of parameters than a time-dependent degradation model. Here, the model parameters include: 1) the decay parameter $\lambda$, 2) the degradation propagation vector $\bm{v}$ which determines the speed and direction of propagation, 3) the parameters, $\rho_1$ and $\rho_2$ in (\ref{eq:R}), that control the standard deviations of the convolution kernel, 4) the parameter $\bm{\theta}$ in the spatial covariance function $c(\cdot; \bm{\theta})$, and 5) the parameter $\bm{\beta}$, a row vector of length $k$ in (\ref{eq:glinear}), that determines the effects of covariates on the degradation generation process. Note that, the parameters, $\lambda$, $\rho_1$, $\rho_2$ and $\bm{v}$, determine the convolution kernel as well as the design matrix $\bm{X}$ in the linear model (\ref{eq:YLinear}). These parameters, together with $\bm{\theta}$, determine the covariance matrix of the error term $\bm{e}$ in (\ref{eq:YLinear}). The last set of parameter $\bm{\beta}$ contains the coefficients of the linear model (\ref{eq:YLinear}).

Although the number of parameters is large, the special structure of the proposed model allows us to estimate the unknown parameters all at once using the Maximum Likelihood Estimation (MLE). Since $\varepsilon_{\Delta}$ is a white-in-time Gaussian spatial process,
\begin{equation}
\tilde{\bm{Y}}(\bm{s},t) = {\bm{Y}}(\bm{s},t) - {\bm{Z}}(\bm{s},t) =  g_{\Delta}(\bm{s},t) + \varepsilon_{\Delta}(\bm{s},t)
\label{eq:diff}
\end{equation}
is also a white-in-time Gaussian spatial process with mean $g_{\Delta}(\bm{s},t)=\bm{x}_0(\bm{s},t) \bm{\beta}^{\top}$ and covariance matrix $\bm{\Sigma}_\varepsilon$. Here, the covariance matrix is determined through the covariance function $c_{\Delta}(\cdot)=\Delta \cdot c(\cdot)$, where $c(\cdot)$ is a covariance function of a white-in-time Gaussian process.

Let $\bm{\Omega}$ be the set that contains the model parameters, the log-likelihood function of $\bm{\Omega}$ given the observations $\tilde{\bm{Y}}(\bm{s},t)$, for $t=2,...,N_t$, are
\begin{equation}
\mathcal{L}(\bm{\Omega}) = \sum_{t=2}^{N_t}\left\{ \frac{1}{ (2\pi)^{N_s/2} (\mathrm{det}\bm{\Sigma}_\varepsilon)^{1/2}}  \exp \left( -\frac{(\tilde{\bm{y}}(\bm{s},t)-\bm{x}_0(\bm{s},t) \bm{\beta}^{\top})^{\top} \bm{\Sigma}_\varepsilon^{-1} (\tilde{\bm{y}}(\bm{s},t)-\bm{x}_0(\bm{s},t) \bm{\beta}^{\top}) }{2} \right) \right \},
\label{eq:likelihood}
\end{equation}
and the Maximum Likelihood (ML) estimator, $\hat{\bm{\Omega}}$, is found by maximizing the log-likelihood function (\ref{eq:likelihood}). It is also possible to obtain the asymptotic confidence intervals for the ML estimators. Under mild regularity conditions, a consistent estimator of the covariance matrix is given by the inverse of the observed information matrix (Meeker and Escobar 1998).

Estimating the unknown parameters of a spatio-temporal process can be computationally costly. Based on the linear model (\ref{eq:glinear}), $\bm{Y}$ is a spatio-temporal process with its covariance structure fully characterized by (\ref{eq:cov}). Hence, one might construct the likelihood function of the model parameter $\bm{\Omega}$ based on the observed $\bm{y}$, which is obtained from the entire spatial and time domains. However, this is not at all practical for large data sets due to an $\mathcal{O}((N_t \times N_s)^3)$ cost for inverting the large covariance matrix (for example, $N_t=20$ and $N_s=441$ in the numerical example). Also note that, for each time point, the convolution on a two-dimensional surface requires $\mathcal{O}(N_s^2)$ operations.

Hence, we leverage the special structure given in (\ref{eq:Y}) and (\ref{eq:G}), and obtain the likelihood function based on $\tilde{\bm{Y}}(\bm{s},t)$ defined in (\ref{eq:diff}), which is a white-in-time Gaussian spatial process. Following this approach, the computational cost is significantly reduced to $\mathcal{O}( N_t \times N_s^3)$, making the proposed model more practical. In fact, both the convolution operation on the two-dimensional space and the computation of the contribution to the total likelihood from the data collected at each time point can be \emph{embarrassingly} parallelized with little effort needed to separate the problem into a number of parallel tasks.

In a special case when the values of $\lambda$, $\rho_1$, $\rho_2$ and $\bm{v}$ are known from some physical knowledge, the parameters, $\bm{\beta}$ and $\bm{\theta}$ associated with a linear model (\ref{eq:glinear}), can also be estimated using the Iteratively Re-Weighted Generalized Least Squares (IRWGLS) which consists of the following steps:

\begin{description}
	\item[Step 1]:
	Set the initial $\hat{\bm{\Sigma}}_{\bm{Y}}$ to an identify matrix of size $N_s \times N_t$.
	\item[Step 2]:
	Estimate $\bm{\beta}$ using the Feasible General Least Squares (FGLS):
	\begin{equation*}
	\hat{\bm{\beta}} = (\bm{X}^{\top} \hat{\bm{\Sigma}}_{\bm{Y}}^{-1} \bm{X})^{-1} \bm{X}^{\top} \hat{\bm{\Sigma}}_{\bm{Y}} \bm{Y}.
	\end{equation*}
	\item[Step 3]:  Based on $ \bm{r}= \bm{Y} - \bm{X} \hat{\bm{\beta}} = (r(s_1,t_1),\ldots,r(s_{N_s},t_{N_t}))^{\top}$, estimate $\hat{\bm{\theta}}$ of the parameter $\bm{\theta}$ associated with the covariance function $c(\cdot; \bm{\theta})$, and obtain the estimate of the covariance matrix, $\hat{\bm{\Sigma}}_{\bm{Y}}$, from equation (\ref{eq:cov}).
	\item[Step 4]: Iterate Steps 1 and 2 until the relative changes of $\hat{\bm{\beta}}$ and $\hat{\bm{\theta}}$ are small.
\end{description}

In the first iteration, since $\hat{\bm{\Sigma}}_{\bm{Y}}$ is an identify matrix, $\hat{\bm{\beta}}$ in Step 1 is the Ordinary Least Squares (OLS) estimator and is unbiased. In subsequent iterations, the finite-sample properties of the FGLS estimator, $\hat{\bm{\beta}}$, are usually unknown and can be studied case-by-case via Monte Carlo experiments. Asymptotically, the FGLS estimator possesses the asymptotic properties of the Maximum Likelihood estimator, and is equivalent to the Generalized Least Squares (GLS) estimator under regularity conditions (Schabenberger and Gotway 2005). Note that, step 3 is computationally expensive and requires an $\mathcal{O}((N_t \times N_s)^3)$ operation if the MLE is used. Hence, one may adopt the fast cross-validation-type method described in Carroll et al. (1997).
Specifically, let $\bm{r}_{(-\bm{s}_0,-t_0)}$ be a column vector of residuals with the residual at location $\bm{s}_0$ and time $t_0$ removed, and let $\eta(\bm{s}_0,t_0)$ be the leaving-one-station-out prediction error, i.e., error in predicting the residual $r(\bm{s}_0,t_0)$ using only $\bm{r}_{(-\bm{s}_0,-t_0)}$ given by
\begin{eqnarray*}
	\eta(\bm{s}_0,t_0)=r(\bm{s}_0,t_0)-\hat{r}(\bm{s}_0,t_0),
\end{eqnarray*}
where $\hat{r}(\bm{s}_0,t_0)$, given below, is the well-defined Simple Kriging predictor which is known to be the best linear estimator of $r(\bm{s}_0,t_0)$ under squared-error loss:
\begin{eqnarray*}
	\hat{r}(\bm{s}_0,t_0)= \bm{\gamma} \bm{\Sigma}_{\bm{r}_{(-\bm{s}_0,-t_0)}}^{-1} \bm{r}_{(-\bm{s}_0,-t_0)}.
\end{eqnarray*}
Here, $\bm{\gamma}=\text{cov}(\bm{r}_{(-\bm{s}_0,-t_0)}^{\top}, r(\bm{s}_0,t_0))$, and the covariance matrix of $\bm{r}_{(-\bm{s}_0,-t_0)}$, $\bm{\Sigma}_{\bm{r}_{(-\bm{s}_0,-t_0)}}$, can be calculated using equation (\ref{eq:cov}). 

The optimum value of $\bm{\theta}$ is found by minimizing the sum of squared leaving-one-station-out prediction errors, i.e.,
\begin{eqnarray}
\min_{\bm{\theta}} \sum_{i}^{N_s}\sum_{j}^{N_t} \eta^2(\bm{s}_i,t_j; \bm{\theta}).
\label{eq:21}
\end{eqnarray}

In summary, the MLE described in Section \ref{sec:MLE} is preferred as it is computationally efficient and the statistical properties of the ML estimator is well studied.

\section{Numerical Results and Model Validation} \label{sec:numerical}
The motivating example presented in Section \ref{sec:degradationst} is re-visited to demonstrate the modeling of spatio-temporal degradation data using the proposed approach. 

\subsection{Estimated Model Parameters}
We present the numerical results and validate the model in this section. Before the MLE is performed, one needs to firstly choose the spatial covariance function $c(\cdot)$ of the process $\varepsilon$ in (\ref{eq:G}). In this numerical example, three candidate covariance functions $c(\cdot)$, which are commonly used for stationary spatial Gaussian process, are assumed and the most appropriate covariance function is then selected based on the model validation. The three spatial covariance functions considered are as follows (Note that, the covariance function $c(\cdot)$ is not the covariance function of the spatio-temporal degradation process which is given in (\ref{eq:cov})):
\begin{itemize}
	\item The Exponential covariance function:
	\begin{eqnarray}
	c(d) = \theta_1 \exp(-d/\theta_2),
	\end{eqnarray}
	where $d$ is the spatial distance, and $\theta_1$ and $\theta_2$ respectively explain the rate of delay and the scale of the spatial correlation.
	\item The Gaussian covariance function:
	\begin{eqnarray}
	c(d) = \theta_1 \exp(-d^2/\theta_2),
	\end{eqnarray}
	where $d$ is the spatial distance, and $\theta_1$ and $\theta_2$ respectively explain the rate of delay and the scale of the spatial correlation.
	\item The Mat\'{e}rn covariance function:
	\begin{eqnarray}
	c(d)=\frac{\theta_1}{2^{\theta_3-1}\Gamma(\theta_3)}(2\theta_3^{1/2}d/\theta_2)^{\theta_3}\mathfrak{K}_{\theta_3}(2\theta_3^{1/2}d/\theta_2)
	\end{eqnarray}
	where $\mathfrak{K}_{\theta_3}$ is a modified Bessel function, the parameters $\theta_1$ and $\theta_2$ are respectively known as the sill and range parameters in spatial statistics, and $\theta_3$ measures the degree of smoothness of the process $\varepsilon$. The Mat\'{e}rn covariance function incorporates the Exponential covariance function as its special case when $\theta_3=1/2$.
\end{itemize}

For this particular numerical example, it is known that the degradation generation within a time interval is proportional to one known covariate which is the pressure applied on the surface. Hence, the length of the parameter $\bm{\beta}$ in (\ref{eq:glinear}) becomes one. Table \ref{table:1} shows the ML estimates assuming different $c(\cdot)$. It is seen from the table that all three models successfully capture the degradation propagation along the south-to-north direction, as the estimated first component of $\bm{v}$ (i.e., the horizontal component) is extremely small. Note that, $\rho_1$ and $\rho_2$ in (\ref{eq:sigma}), respectively control the standard deviations of the convolution kernel $\omega_\Delta$ in directions which are parallel and perpendicular to the direction of degradation propagation. We see, from Table \ref{table:1}, that the estimated value of $\rho_1$ is also greater than $\rho_2$, indicating a higher level of uncertainty along the main degradation propagation direction. The estimated effect of degradation generation, $\beta$, is close to 1 from all three models. Before further analysis on the estimation results is possible, model validation is needed to select the most appropriate parametric form for the covariance function $c(\cdot)$.

\begin{table}
	\centering
	\caption{The estimated model parameters assuming different $c(\cdot)$}
	\label{table:1}
	\begin{tabular}{cc|c|c|c|c}
		\cline{3-5}
		& & \multicolumn{3}{ c| }{covariance function $c(\cdot)$} \\ \cline{3-5}
		& & Exponential & Gaussian & Mat\'{e}rn  \\ \cline{1-5}
		\multicolumn{1}{ |c  }{\multirow{2}{*}{ parameters} } &
		\multicolumn{1}{ |c| }{ $\lambda$ } & 0.127 & 0.09 & 0.149 &     \\ \cline{2-5}
		\multicolumn{1}{ |c  }{}                        &
		\multicolumn{1}{ |c| }{$\bm{v}$} & (-0.040,0.499) & (-0.004,0.793) & (0.006,0.598) &     \\ \cline{2-5}
		\multicolumn{1}{ |c  }{}                        &
		\multicolumn{1}{ |c| }{$\rho_1$} & 1.119 & 2.247 & 0.802 &     \\ \cline{2-5}
		\multicolumn{1}{ |c  }{}                        &
		\multicolumn{1}{ |c| }{$\rho_2$} & 0.192 & 0.301 & 0.216 &     \\ \cline{2-5}
		\multicolumn{1}{ |c  }{}                        &
		\multicolumn{1}{ |c| }{$\theta_1$} & 0.019 & 0.010 & 0.071 &     \\ \cline{2-5}
		\multicolumn{1}{ |c  }{}                        &
		\multicolumn{1}{ |c| }{$\theta_2$} & 12.883 & 11.564 & 81.560 &     \\ \cline{2-5}
		\multicolumn{1}{ |c  }{}                        &
		\multicolumn{1}{ |c| }{$\theta_3$} & N/A & N/A & 0.434 &     \\ \cline{2-5}
		\multicolumn{1}{ |c  }{}                        &
		\multicolumn{1}{ |c| }{ $\beta$ } & 0.977 & 1.251 & 1.108 &     \\ \cline{1-5}
	\end{tabular}
\end{table}

\subsection{Model Validation and Selection}
To validate the model and select the best $c(\cdot)$, we again resort to the key observation that $\tilde{\bm{Y}}(\bm{s},t)$ is a white-in-time Gaussian spatial process with mean $g_{\Delta}(\bm{s},t)=\bm{x}_0(\bm{s},t) \bm{\beta}^{\top}$ and covariance matrix $\bm{\Sigma}_\varepsilon$. Once the model parameters have been estimated, the residuals, $\bm{y}^*(\bm{s},t)=\tilde{\bm{y}}(\bm{s},t)-\bm{x}_0(\bm{s},t) \hat{\bm{\beta}}^{\top}$ for $t=1,...,N_t$, should look like a number of $N_t$ samples generated from a white-in-time Gaussian spatial process $\varepsilon_{\Delta}$ with the covariance function $c_{\Delta}(\cdot)=\Delta \cdot c(\cdot)$. Hence, the model can be validated \emph{graphically} by comparing the empirical semi-variogram estimated from the residuals $\bm{y}^*(\bm{s},t)$ and the theoretical semi-variogram computed based on the estimated model parameters.
This simple but elegant idea for model validation is similar to that of choosing the best probability distribution for lifetime data using probability plotting, which has been widely adopted in statistical reliability analysis (Meeker and Escobar 1998).
Also note that, this approach validates not only the choice of the covariance function $c(\cdot)$, but also the fitted degradation generation and propagation processes that yield the residuals.

To estimate the empirical semi-variogram of the residuals $\bm{y}^*(\bm{s},t)$, the well-known Cressie-Hawkins robust estimator is used (Cressie and Hawkins 1980). Since the residual, $\bm{Y}^*$, is a white-in-time process, we may write, for our problem, the Cressie-Hawkins estimator as follows:
\begin{eqnarray}
\hat{\gamma}(d) = \left\{ \frac{1}{|N(d)|} \sum_{N(d)} \sum_{t=2}^{N_t}|\bm{y}^*(\bm{s}_i,t)-\bm{y}^*(\bm{s}_j,t)|^{1/2}   \right\}^4 \times \left( 0.914 + \frac{0.988}{|N(d)|} \right)
\label{eq:Cressie}
\end{eqnarray}
where $N(d)=\{(i,j): |\bm{s}_i-\bm{s}_j| =d  \}$ and $|N(d)|$ is the number of distinct elements in the set $N(d)$.

\begin{figure}[!htb]
	\begin{center}
		\includegraphics[width=1\textwidth]{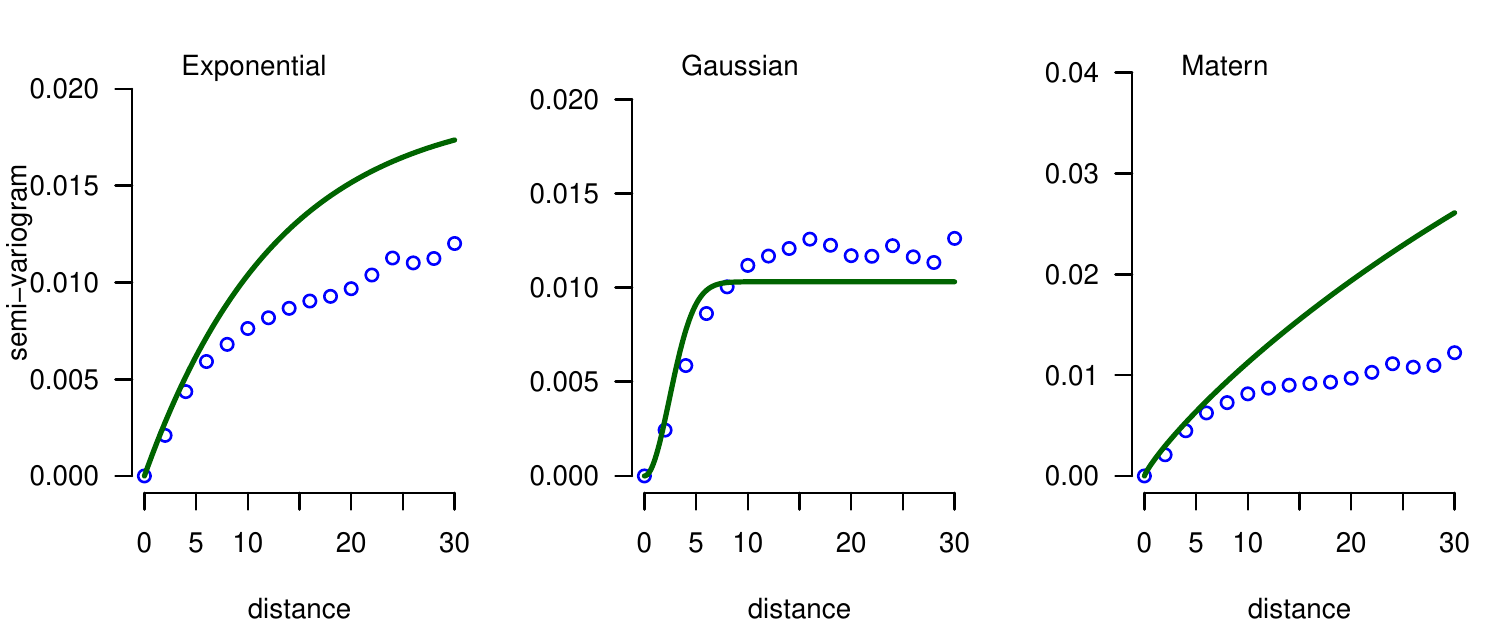}
		\caption{Comparison between the empirical semi-variogram estimated from the Cressie-Hawkins estimator and the theoretical semi-variogram computed based on the estimated model parameters, respectively assuming Exponential, Gaussian, and Mat\'{e}rn covariance functions for $c(\cdot)$.} \label{fig:variogram}
	\end{center}
\end{figure}
Figure \ref{fig:variogram} shows the comparison between the empirical semi-variogram and the theoretical semi-variogram computed based on the estimated model parameters, respectively assuming Exponential, Gaussian, and Mat\'{e}rn covariance functions. It is immediately seen that the best match between the empirical and the theoretical variogram is obtained if the Gaussian covariance function is chosen for $c(\cdot)$.
\begin{figure}[!htb]
	\begin{center}
		\includegraphics[width=1\textwidth]{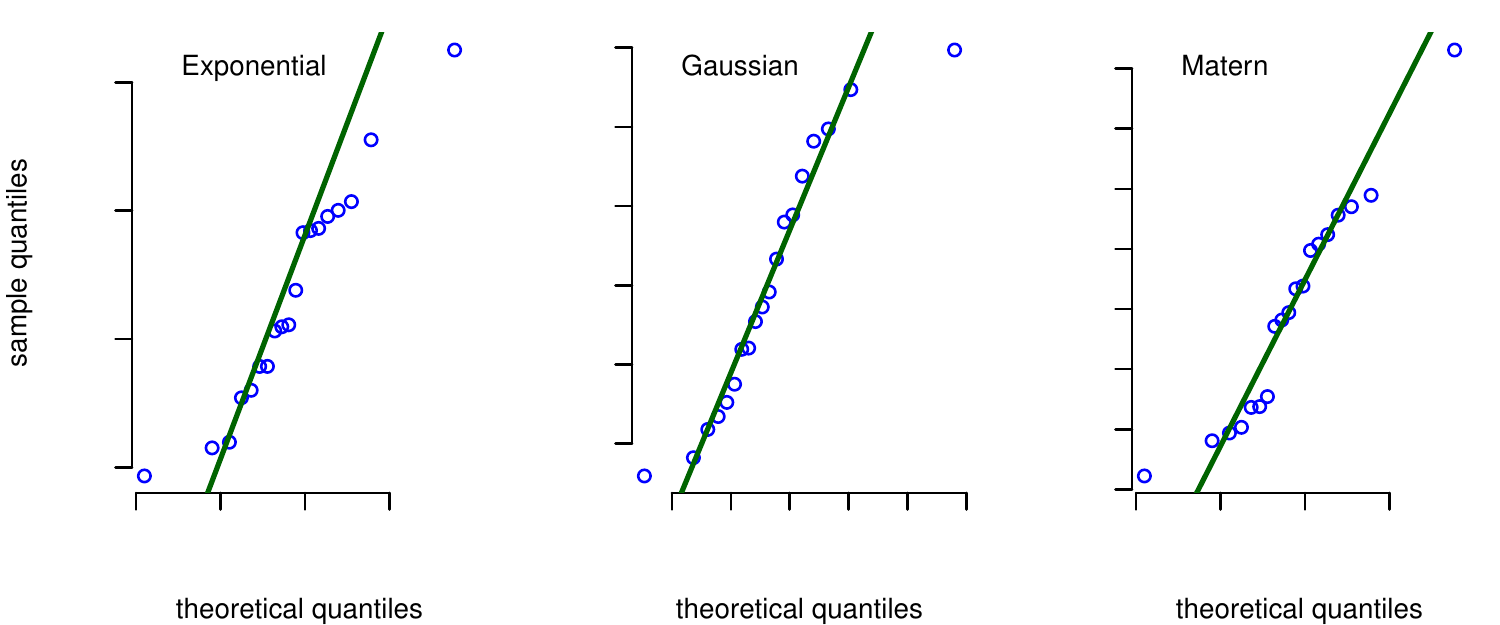}
		\caption{Chi-square q-q plot to check the multivariate normality of the residuals, respectively assuming Exponential, Gaussian, and Mat\'{e}rn covariance functions for $c(\cdot)$.} \label{fig:qqplot}
	\end{center}
\end{figure}

The general normality check of the residual process can be performed using the conventional chi-square q-q plot, and the plot is shown in Figure \ref{fig:qqplot}. Again, when the Gaussian covariance function is chosen for $c(\cdot)$, we see the best fit between the theoretical and sample quantiles of the residuals $\tilde{\bm{y}}(\bm{s},t)$.

\begin{table}
	\centering
	\caption{ML estimates with Gaussian covariance function $c(\cdot)$}
	\label{table:2}
	\begin{tabular}{ p{2cm}||p{2.5cm}|p{2cm}|p{3.5cm} }
		\hline
		Parameter & ML Estimate & Standard Error & Approximate 90\% Confidence Interval\\
		\hline
		$\lambda$  & 0.09    & 0.012 &    (0.071,0.109)\\
		$v_1$  &   -0.004 & 0.018 &(-0.033,0.025) \\
		$v_2$ &0.793 & 0.039 &  (0.728,0.857)\\
		$\rho_1$     &2.247 & 0.117&  (2.054,2.440)\\
		$\rho_2$  &   0.301  & 0.024& (0.262,0.339)\\
		$\theta_1$ & 0.010 & 0.0008  &(0.009,0.012)\\
		$\theta_2$ & 11.564 & 0.523&(10.703,12.425)\\
		$\beta$ & 1.251 & 0.040&(1.184,1.317)\\
		\hline
	\end{tabular}
\end{table}

\begin{figure}[!htb]
	\begin{center}
		\includegraphics[width=1\textwidth]{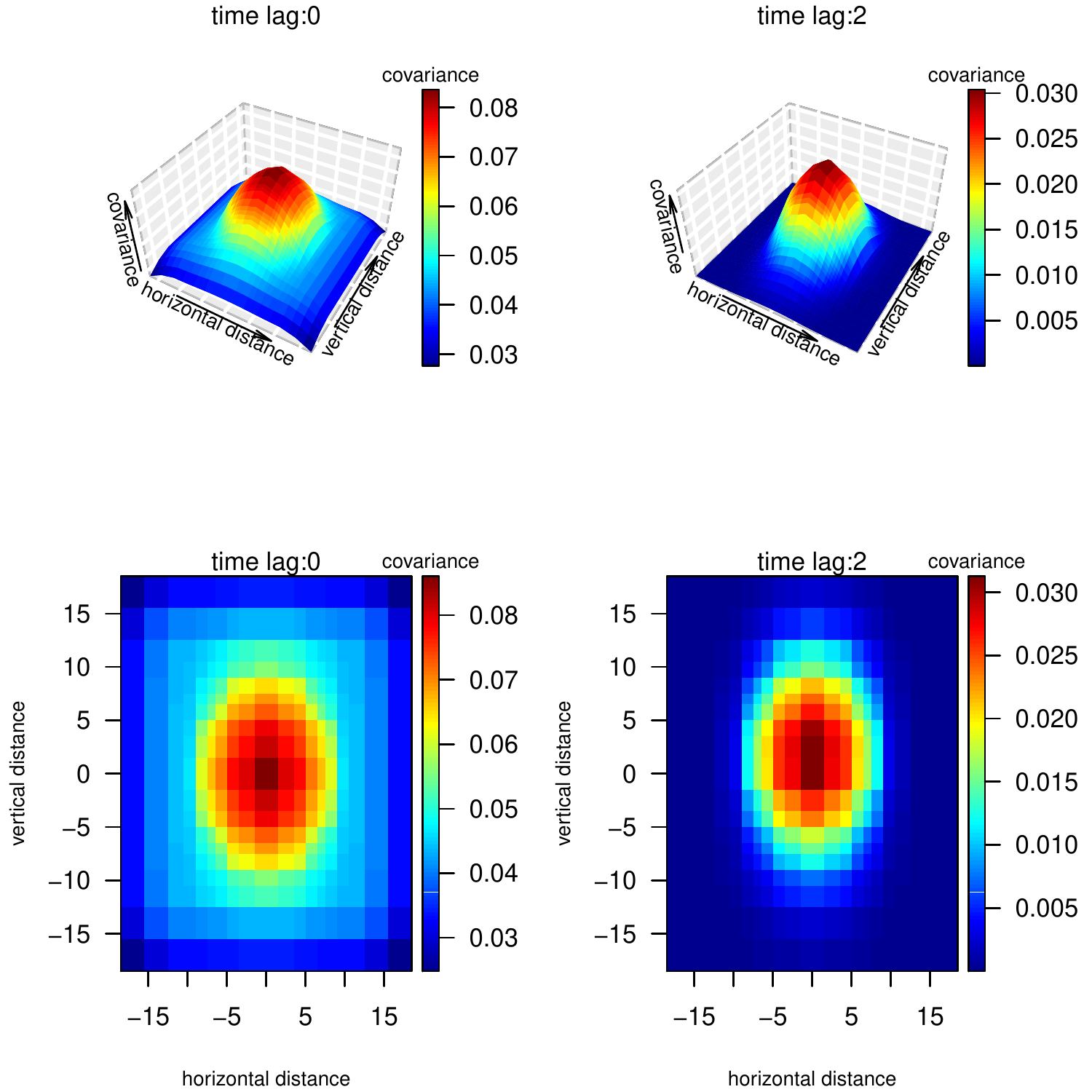}
		\caption{Covariance functions for different spatial and time lags. The two plots on the left respectively show the surface and contour plots of the covariance function for time lag 0, while the two plots on the right respectively show the surface and contour plots of the covariance function for time lag 2.} \label{fig:cov}
	\end{center}
\end{figure}

Based on the results shown in Table \ref{table:1} and the Gaussian covariance function chosen for $c(\cdot)$, a summary of the final ML estimates are given in Table 2. It can be seen from Table 2 that the amount of degradation propagated to its neighboring areas decays to 50\% of its original value approximately after 8 time periods, given that $\hat{\lambda}=0.09$. The spatial propagation of degradation follows the south-to-north direction, and the propagation speed is about 0.793 per unit time based on the estimated $\bm{v}$. In addition, a much higher uncertainty is expected in the north-south direction than in the west-east direction, given that $\hat{\rho}_1$ is 7 times higher than $\hat{\rho}_2$. For the white-in-time Gaussian spatial process $\varepsilon_{\Delta}$, the sill (i.e., the value at which the semi-variagram levels off) of its estimated semi-variogram is 0.01 based on $\hat{\theta}_1$, and the practical range (i.e., the spatial lag distance at which the semi-variogram reaches 95\% of the sill) is close to 6. Since the degradation data are aggregated to a $21 \times 21$ pixel array, the spatial process $\varepsilon_{\Delta}$ has a rather local effect on $G_{\Delta}(\bm{s},t)$ in (\ref{eq:G}). The covariance function of the spatio-temporal degradation process $Y(\bm{s},t)$ can now be calculated by (\ref{eq:cov}). Figure \ref{fig:cov} shows the computed covariance for different spatial and time lags. In particular, the two plots on the left respectively show the surface and contour plots of the covariance function for time lag 0, while the two plots on the right respectively show the surface and contour plots of the covariance function for time lag 2. It is seen that the covariance decreases as the time lag increases, as expected. For a fixed time lag, in particular, the covariance decreases faster in the horizontal direction as the degradation propagates vertically in this example.

\begin{figure}[!htb]
	\begin{center}
		\includegraphics[width=1\textwidth]{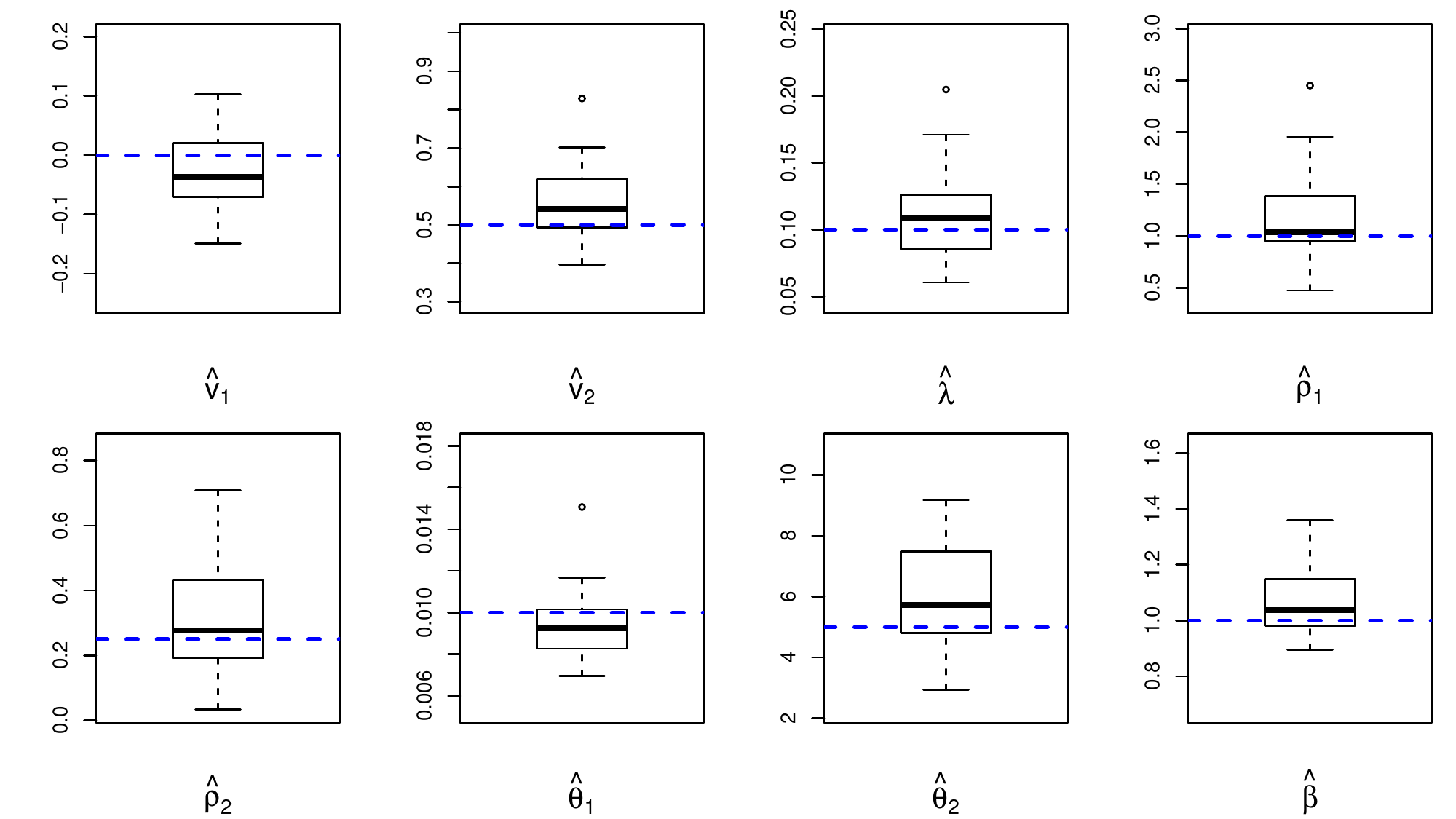}
		\caption{Boxplot of the ML estimates obtained from 500 simulated degradation data set. For each parameter, the ``true'' value which is used in generating the simulation data is also indicated by the dashed horizontal line. } \label{fig:boxplot}
	\end{center}
\end{figure}

Simulation studies are also used to investigate the performance of the ML estimators. A number of 500 degradation data sets are generated based on the following specified values: $\lambda=0.1$, $v_1=0$, $v_2=0.5$, $\rho_1=1$, $\rho_2=0.25$, $\theta_1=0.01$, $\theta_2=5$ and $\beta=1$ (the simulation procedure is described in Section \ref{sec:FPLT}). For each simulated data set, the ML estimates of the model parameters are obtained. Figure \ref{fig:boxplot} shows the boxplot of the ML estimates, for all model parameters, obtained from the simulated data sets. For each parameter, the ``true'' value which is used to simulate the data is also indicated by the dashed horizontal line. We see from the simulation results that the ML estimation is an effective approach in estimating the unknown model parameters. For example, the bias of the ML estimators is relatively small, and the mean-squared-error of the estimates are calculated as, $4.96 \times 10^{-3}$, $1.18 \times 10^{-2}$, $1.11 \times 10^{-3}$, $2.41 \times 10^{-1}$, $2.76 \times 10^{-2}$, $2.86 \times 10^{-6}$, $3.58$ and $1.88 \times 10^{-2}$, which are also small given the scale of the ``true'' parameters values.

\section{Additional Applications: First-Passage-Time and First-Passage-Location} \label{sec:FPLT}
In degradation data analysis, the First-Passage-Time (FPT) of the degradation process for some given threshold is of great importance. The FPT is usually used to approximate the time-to-failure for reliability prediction, or, to determine the time for preventive maintenance (Liu et al. 2012). In the literature, the FPT for a pure time-dependent stochastic degradation process defined in (\ref{eq:degradation}) is given by
\begin{eqnarray}
T^* = \inf\{t: Y(t) \geq y^* \}.
\label{eq:FPT1}
\end{eqnarray}
where $y^*$ is the pre-specified threshold. In general, the closed-form expression of the distribution of $T^*$ is not available. In some special cases when the degradation process is modeled by a Wiener process, it is well-known that the FPT follows the Inverse Gaussian distribution.

For spatio-temporal degradation process, one might be interested in not only when the degradation process hits the pre-determined threshold but also where it hits the threshold. In particular, the FPT for a spatio-temporal degradation process can be defined as:
\begin{eqnarray}
T^* = \inf\{t: \max_{\bm{s}}(Y(\bm{s}, t)) \geq y^* \}.
\label{eq:FPT2}
\end{eqnarray}

Numerical simulation is needed to approximate the distribution of FPT. The simulation of random fields has been discussed in the literature (Lang and Potthoff 2011, Brouste et al., 2007, Sigrist et al., 2015). For the spatio-temporal model described in this paper, the degradation process can be directly simulated using equations (\ref{eq:Y}) and (\ref{eq:G}) in an iterative manner. This idea is similar to Sigrist et al. (2015) in which the authors investigate the generation of a Gaussian random field from stochastic PDE. In particular, at any time $t$:

\begin{itemize}
	\item compute the amount of degradation, $g_{\Delta}(\bm{s},t)$, generated over the time interval $(t-\Delta,t]$ in equation (\ref{eq:G}).
	\item compute the propagation of degradation, $Z(\bm{s},t)$, using the first line of equation (\ref{eq:Z}). This involves a convolution operation on a two-dimensional space.
	\item simulate the noise term of equation (\ref{eq:G}), $\varepsilon_{\Delta}(\bm{s},t)$.
	\item the simulated degradation at time $t$ is computed by equations (\ref{eq:Y}) and (\ref{eq:G}), i.e., the sum of the three terms computed in steps 1 to 3:
	\begin{equation}
	Y(\bm{s},t) = g_{\Delta}(\bm{s},t) + Z(\bm{s},t) + \varepsilon_{\Delta}(\bm{s},t).
	\nonumber
	\end{equation}
\end{itemize}

\begin{figure}[!htb]
	\begin{center}
		\includegraphics[width=0.7\textwidth]{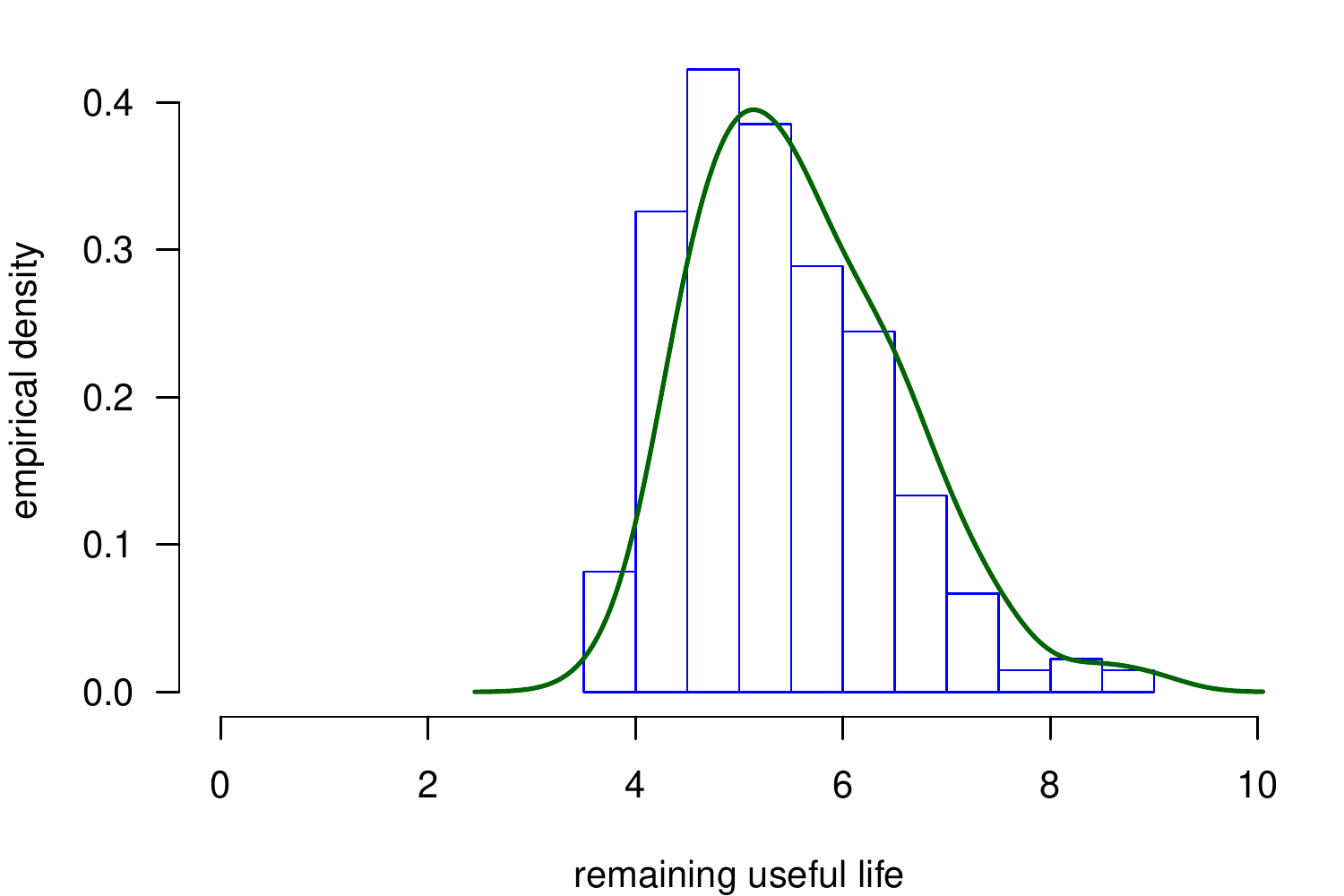}
		\caption{The histogram and the kernel density of the simulated remaining useful life.} \label{fig:FPT}
	\end{center}
\end{figure}

It is easy to see that the computational complexity of the simulation procedure described above is dominated by simulating the spatial noise term $\varepsilon_{\Delta}(\bm{s},t)$ which typically requires $\mathcal{O}( N_t \times N_s^3)$ operations. For illustrative purposes, we arbitrarily choose a threshold value which is 70\% higher than the highest degradation level at time $t=20$, and simulate the degradation data from 500 simulation runs. Figure \ref{fig:FPT} shows the histogram as well as the density estimated from the kernel density estimation for the simulated Remaining Useful Life (RUL).
It is seen that there is a high probability that the degradation level will hit the threshold in the following 4 to 8 time periods.

\begin{figure}[!htb]
	\begin{center}
		\includegraphics[width=0.8\textwidth]{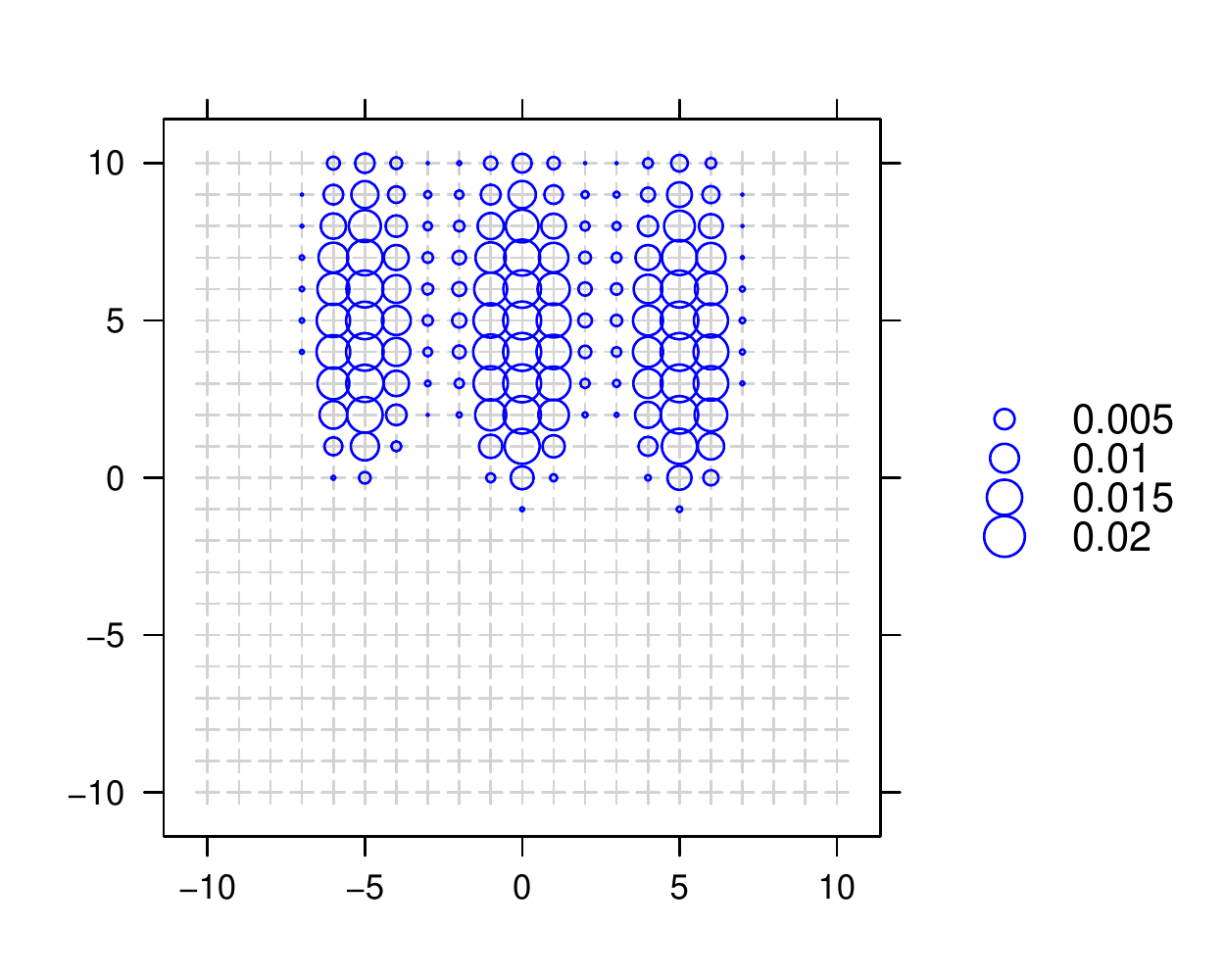}
		\caption{The spatial distribution of the simulated FPL. The radius of each circle is proportional to the probability (shown by the legend on the right) that the degradation will hit the threshold at the location where the circle resides.} \label{fig:FPL}
	\end{center}
\end{figure}

For each simulation run, the location where the spatial degradation hits the threshold for the first time is also obtained. Figure \ref{fig:FPL} shows the spatial distribution of the simulated First-Passage-Location (FPL). In this figure, the radius of each circle is proportional to the probability that the degradation will hit the threshold at the location where the circle resides. Since three regions with high degradation levels are observed in Figure \ref{fig:example1}, it is not surprising at all to note from Figure \ref{fig:FPL} that the FPL is most likely to be found in one of these three regions.

\section{Conclusions} \label{sec:conclusion}

This paper proposed a statistical degradation model for spatio-temporal degradation data. The proposed approach models the degradation process by a random field with a space-time non-separable and anisotropic covariance structure. The degradation at a particular location and time is expressed as the sum of a spatial degradation generation process, and a spatio-temporal degradation propagation process based on the convolution operation. Some unique challenges associated with the modeling of spatio-temporal degradation data were discussed, and a numerical example was presented to demonstrate the application of the proposed approach, including the parameter estimation, model validation and the approximations of the distribution of FPT and FPL using simulation. Note that, it has been assumed in this paper that the propagation of degradation is uniform in space and time-invariant, i.e., the degradation propagates along the same direction and at a constant speed. In some applications, it is possible that the propagation of degradation is affected by dynamic environmental conditions. In recent years, the modeling of pure time-dependent degradation data under dynamic environments has received much attention (Liao and Tian 2012, Zhou, Serban and Gebraeel 2014, Bian et al., 2015, Hong et al., 2015). Hence, extending the current modeling framework so as to incorporate dynamic environments into the spatio-temporal degradation model is certainly an important topic and worth further investigating in the further research.

\section{Appendix A: Approximation Error}  \label{sec:appendix}
Since equation (14) is used to approximate equation (12) when  $n$ is large, the error of approximation is obtained by taking the difference of these two equations:
\begin{eqnarray}
\begin{split}
e_n  = & \sum_{i=n+1}^{\infty} e^{-i\lambda\delta}\phi_{i} \left( \bm{s} \right)* \left(g_{\delta}(\bm{s},t-i\delta)+\varepsilon_{\delta}(\bm{s},t-i\delta) \right)    \\ & + e^{-n\lambda\delta}\phi_{n}\left( \bm{s} \right) *  Z(\bm{s},t-n\delta).
\label{eq:error}
\end{split}
\end{eqnarray}

Hence, $e_n$ can be ignored when $n$ is large. Because $g$ is deterministic and $\varepsilon$ is the white-in-time spatially colored Gaussian noise, for any $i \geq n+1$ and $\epsilon>0$, there must exist a finite $M$ such that
\begin{eqnarray}
\begin{split}
& P\left\{ \left |  \frac{e^{-i\lambda\delta}\phi_{i}(\bm{s})\left(g_{\delta}(\bm{s},t-i\delta)+\varepsilon_{\delta}(\bm{s},t-i\delta) \right) }{e^{-n\lambda\delta}}  \right |>M \right\} \\ & = P\left\{ \left | e^{-(i-n)\lambda\delta}\phi_{i}(\bm{s})\left(g_{\delta}(\bm{s},t-i\delta)+\varepsilon_{\delta}(\bm{s},t-i\delta) \right)   \right |>M \right\} < \epsilon,
\label{eq:error1}
\end{split}
\end{eqnarray}
and
\begin{eqnarray}
\begin{split}
& P\left\{ \left |  \frac{e^{-n\lambda\delta}\phi_{i}(\bm{s}) Z(\bm{s},t-n\delta)}{e^{-n\lambda\delta}}  \right |>M \right\} \\ & = P\left\{  \left | \phi_{i}(\bm{s}) Z(\bm{s},t-n\delta)    \right| >M  \right\} < \epsilon.
\label{eq:error2}
\end{split}
\end{eqnarray}
Hence, every summation term for $e_n$ is $\mathcal{O}_p(e^{-n\lambda\delta})$ and so is $e_n$.

\section{Appendix B: Derivation of the Covariance (\ref{eq:cov})}  \label{sec:appendixB}

We show how equation (\ref{eq:cov}) is derived. It follows from (\ref{eq:Y1}) that the covariance, $\mathrm{cov}(Y(\bm{s}_1,t_1),Y(\bm{s}_2,t_2))$ for $t_1 \leq t_2$, can be written as:
\begin{eqnarray}
\begin{split}
& \mathrm{cov}(Y(\bm{s}_1,t_1),Y(\bm{s}_2,t_2)) = \\ &  \quad \mathrm{cov} ( \sum_{i=1}^{\infty}(\Psi_{i}(\bm{s}_1)*\varepsilon_{\delta}(\bm{s}_1,t_1-i\delta)), \varepsilon_{\delta}(\bm{s}_2,t_2))\\
& \quad + \mathrm{cov} ( \varepsilon_{\delta}(\bm{s}_1,t_1), \varepsilon_{\delta}(\bm{s}_2,t_2))\\
& \quad + \mathrm{cov} ( \sum_{i=1}^{\infty}(\Psi_{i}(\bm{s}_2)*\varepsilon_{\delta}(\bm{s}_2,t_2-i\delta)), \varepsilon_{\delta}(\bm{s}_1,t_1)) )\\
& \quad + \mathrm{cov} ( \sum_{i=1}^{\infty}(\Psi_{i}(\bm{s}_1)*\varepsilon_{\delta}(\bm{s}_1,t_1-i\delta)), \\ & \quad \quad \quad \quad \quad \sum_{i=1}^{\infty}(\Psi_{i}(\bm{s}_2)*\varepsilon_{\delta}(\bm{s}_2,t_2-i\delta))).
\label{eq:a1}
\end{split}
\end{eqnarray}

In what follows, we obtain the expression for each term on the right hand side (RHS) of (\ref{eq:a1}).
Since $\varepsilon_{\delta}(\bm{s},t)$ is a white-in-time isotropic random field with spatial covariance function $c_{\delta}(\cdot)$, we immediately obtain the expressions of the first two terms on the RHS of (\ref{eq:a1}):
\begin{eqnarray}
\mathrm{cov} ( \sum_{i=1}^{\infty}(\Psi_{i}(\bm{s}_1)*\varepsilon_{\delta}(\bm{s}_1,t_1-i\delta)), \varepsilon_{\delta}(\bm{s}_2,t_2))=0,  \quad t_1 \leq t_2,
\end{eqnarray}
and
\begin{equation}
\mathrm{cov} ( \varepsilon_{\delta}(\bm{s}_1,t_1), \varepsilon_{\delta}(\bm{s}_2,t_2))=
\begin{cases}
c_{\delta}(\bm{d}), & \text{if}\ t_1 = t_2 \\
0, & \text{otherwise}
\end{cases}
\end{equation}
where $\bm{d}=\bm{s}_2-\bm{s}_1$ is a vector. Note that, since the random field $\varepsilon_{\delta}(\bm{s},t)$ is isotropic, $c_{\delta}(\bm{d})=c_{\delta}(||\bm{d}||)$ with $||\bm{d}||$ representing the distance between $\bm{s}_1$ and $\bm{s}_2$.

Let $t_1=t_2-\jmath \delta$ for some $\jmath \geq 0$, the expression of the third term on the RHS of (\ref{eq:a1}) is derived as follows:
\begin{eqnarray}
\begin{split}
\mathrm{cov} & ( \sum_{i=1}^{\infty}(\Psi_{i}(\bm{s}_2)*\varepsilon_{\delta}(\bm{s}_2,t_2-i\delta)), \varepsilon_{\delta}(\bm{s}_1,t_1))) \\
& = \mathrm{cov} ( \Psi_{\jmath,t_2}(\bm{s}_2)*\varepsilon_{\delta}(\bm{s}_2,t_1), \varepsilon_{\delta}(\bm{s}_1,t_1) )\\
& = \mathrm{cov} ( \int_{\mathbb{R}^2}^{} \Psi_{\jmath,t_2}(\bm{x}) \varepsilon_{\delta}(\bm{s}_2-\bm{x},t_1) d\bm{x}, \varepsilon_{\delta}(\bm{s}_1,t_1)  )\\
& = (\Psi_{\jmath,t_2}*c_{\delta})( \bm{d} ).
\end{split}
\end{eqnarray}

The expression of the fourth term on the RHS of (\ref{eq:a1}) can be derived in a similar way as follows:
\begin{eqnarray}
\begin{split}
\mathrm{cov} & ( \sum_{i=1}^{\infty}(\Psi_{i}(\bm{s}_1)*\varepsilon_{\delta}(\bm{s}_1,t_1-i\delta)), \sum_{i=1}^{\infty}(\Psi_{i}(\bm{s}_2)*\varepsilon_{\delta}(\bm{s}_2,t_2-i\delta))) \\ &= \sum_{i=1}^{\infty} \mathrm{cov}( \Psi_{i}(\bm{s}_1)*\varepsilon_{\delta}(\bm{s}_1,t_1-i\delta)),  \Psi_{\jmath+i,t_2}(\bm{s}_2)*\varepsilon_{\delta}(\bm{s}_2,t_2-\jmath\delta-i\delta)) )\\
& = \sum_{i=1}^{\infty} ( \tilde{\Psi}_{i} * \Psi_{\jmath+i,t_2} * c_{\delta} )(\bm{d})
\label{eq:a2}
\end{split}
\end{eqnarray}
where $\tilde{\Psi}_{i}(\bm{s}) \equiv \Psi_{i}(-\bm{s})$.

Note that, $\tilde{\Psi}_{i}$ and $\Psi_{i}$ are Dirac delta functions when $i=0$, we have
\begin{eqnarray}
\begin{split}
\mathrm{cov}(Y(\bm{s}_1,t_1),Y(\bm{s}_2,t_2)) &  =  (\Psi_{\jmath,t_2}*c_{\delta})( \bm{d} ) + \sum_{i=1}^{\infty} ( \tilde{\Psi}_{i} * \Psi_{\jmath+i,t_2} * c_{\delta} )(\bm{d}) + I_{\{\jmath=0\}}c_{\delta}(\bm{d})\\
& = \sum_{i=0}^{\infty} ( \tilde{\Psi}_{i} * \Psi_{\jmath+i,t_2} * c_{\delta} )(\bm{d}) + I_{\{\jmath=0\}}c_{\delta}(\bm{d}).
\end{split}
\end{eqnarray}
where $I_{\{\jmath=0\}}=1$ only when ${\jmath=0}$, otherwise, $I_{\{\jmath=0\}}=0$.

\section{Appendix C: The Connection to Physical Degradation Models}  \label{sec:appendixC}
The connection, between the proposed spatio-temporal degradation model and physical degradation models, is firstly established under a general setting. Then, we focus on a special case of a real physical degradation model.

Consider a general PDE which consists of four major components including generation, convection, diffusion and decay:
\begin{equation} \label{eqn:transport}
\frac{\partial \varphi(\bm{s},t)}{\partial t} = Q(\bm{s},t) - \nabla \cdot [\bm{v}(\bm{s},t) \varphi(\bm{s},t)] + \nabla \cdot [\bm{K}(\bm{s},t) \cdot \nabla \varphi(\bm{s},t)]  - \frac{1}{\tau}\varphi(\bm{s},t)
\end{equation}
where $\varphi$ is the quantity of interest at location $\bm{s}$ and time $t$, $Q$ is the generation rate, $\bm{v}$ is the propagation velocity vector, $\bm{K}$ is a second-order tensor of diffusivity, and $\tau$ is the relaxation timescale of decay. Note that, the second and the third terms on the RHS of (\ref{eqn:transport}) respectively represent the convection and diffusion processes.

First, we show that the convolution model (\ref{eq:Z}) is an approximation to the physical convection-diffusion process under special conditions. Under a uniform and steady degradation propagation velocity field in a homogeneous space with zero degradation generation, (\ref{eqn:transport}) can be simplified as
\begin{equation}\label{eqn:simple_transport}
\frac{\partial \varphi(\bm{s},t)}{\partial t} = - \bm{v} \cdot \nabla \varphi(\bm{s},t) + K \nabla^2 \varphi(\bm{s},t) - \frac{1}{\tau}\varphi(\bm{s},t).
\end{equation}

Applying the Fourier transform, equation (\ref{eqn:simple_transport}) becomes
\begin{equation} \label{eqn:Fourier_transport}
\frac{d \tilde{\varphi}_\eta(t)}{d t} = - i (\bm{\eta} \cdot \bm{v}) \tilde{\varphi}_\eta(t) - \eta^2 K \tilde{\varphi}_\eta(t) - \frac{1}{\tau} \tilde{\varphi}_\eta(t),
\end{equation}
in which $\eta$ is the wave number and $\tilde{\varphi}_\eta$ is the Fourier coefficient of $\varphi$. It is not difficult to show that the solution of equation (\ref{eqn:Fourier_transport}) is given by
\begin{equation}
\tilde{\varphi}_\eta(t+\Delta) = \exp \left\{ -\frac{1}{\tau} \Delta -[ i (\bm{\eta} \cdot \bm{v}) + K \eta^2] \Delta  \right\} \tilde{\varphi}_\eta(t),
\end{equation}
and the backward Fourier transform leads to
\begin{equation} \label{eqn:conv_transport}
\varphi(\bm{s},t) = \exp\left\{ -\frac{1}{\tau} \Delta \right\} \int_{\mathbb{R}^2} \Omega_\Delta(\bm{x}) \varphi(\bm{s}-\bm{x},t-\Delta) d\bm{x},
\end{equation}
in which the kernel $\Omega_{\Delta}(\bm{x})$ is exactly a Gaussian kernel:
\begin{equation}
\Omega_\Delta(\bm{x}) = \frac{1}{2\pi|\bm{\Sigma}_{\Delta}|^{1/2}}\exp\left\{-\frac{(\bm{x} - \bm{v} \Delta)^{\top} \bm{\Sigma}_{\Delta}^{-1}(\bm{x} - \bm{v} \Delta)}{2}\right\},
\end{equation}
with the covariance matrix given by
\begin{equation} \label{eqn:cov_mat}
\bm{\Sigma}_\Delta =
\begin{pmatrix}
2 K \Delta & 0\\
0 & 2 K \Delta
\end{pmatrix}.
\end{equation}
The covariance matrix (\ref{eqn:cov_mat}) is known as the diffusivity tensor in physics.
The comparison between (\ref{eq:Z1}), (\ref{eq:Z2}) and (\ref{eqn:conv_transport}) immediately justifies why Gaussian convolution kernel is chosen with $\lambda = \tau^{-1}$.

Equation (\ref{eqn:cov_mat}) also suggests that
\begin{equation}
\rho_1 = 2 K_\parallel(s,t), \quad\quad \rho_2 = 2 K_\perp(s,t)
\end{equation}
where $K_\parallel$ and $K_\perp$ are the diffusivities respectively parallel and perpendicular to the propagation direction. In a special case when an isotropic diffusion model is used (i.e., $K = K_\parallel = K_\perp$), the covariance matrix reduces to equation (\ref{eqn:cov_mat}). The connection shown above shows that the proposed statistical model is well motivated when \emph{the direction and speed of degradation propagation are slowly varying within a certain spatial region}.

To see the relationship between $g_{\Delta}(\bm{s},t)$ and the source term $Q(\bm{s},t)$ in (\ref{eqn:transport}), we now consider the limit situation with zero diffusion and no decay, i.e., $\bm{K} \rightarrow 0$ and $\tau \rightarrow \infty$. Then, equation (\ref{eqn:transport}) reduces to
\begin{equation} \label{eqn:convection}
\frac{\partial \varphi(\bm{s},t)}{\partial t} = - \nabla \cdot [\bm{v}(\bm{s},t) \varphi(\bm{s},t)] + Q(\bm{s},t).
\end{equation}

It is obvious that $\zeta_\Delta = 1$ and the convolution kernel $\omega_\Delta$ in (\ref{eq:Z1}) in this limit becomes a Dirac delta function,
\[
\omega_\Delta(\bm{x}) = \delta(\bm{x} - \bm{v} \Delta).
\]
Hence, substituting (\ref{eq:Z}) into (\ref{eq:Y}) and (\ref{eq:G}), and omitting the error process yields
\begin{equation} \label{eqn:conv_proc}
Y(\bm{s},t) = \{ \omega_\Delta * Y(\bm{s},t-\Delta)\} + g_{\Delta}(\bm{s},t) = Y(\bm{s}-\bm{v} \Delta, t - \Delta) + g_{\Delta}(\bm{s},t).
\end{equation}

Since $Y(\cdot,t)$ in the statistical model corresponds to the quantity $\varphi$ at $t$ in the physical model, equation (\ref{eqn:conv_proc}) can be rewritten as
\begin{equation} \label{eqn:conv_proc_1}
\varphi(\bm{s},t) = \varphi(\bm{s}-\bm{v}\Delta, t- \Delta) + g_{\Delta}(\bm{s},t).
\end{equation}

Applying Taylor expansion and assuming the mass conservation (i.e., $\nabla \cdot \bm{v}(\bm{s},t) = 0$), equation (\ref{eqn:conv_proc_1}) yields
\begin{equation} \label{eqn:conv_approx}
\Delta \left( \frac{\partial \varphi(\bm{s},t)}{\partial t} + \nabla \cdot [\bm{v}(\bm{s},t) \varphi(\bm{s},t)] - [g_{\Delta}(\bm{s},t) / \Delta] \right) =  O(\Delta^2).
\end{equation}
By comparing (\ref{eqn:convection}) and (\ref{eqn:conv_approx}), it is clear that the convolution model (\ref{eqn:conv_proc}) is a first-order approximation to the scalar transport equation (\ref{eqn:convection}), i.e., (\ref{eqn:conv_approx}) converges to (\ref{eqn:convection}) as $\Delta \rightarrow 0$.
Hence, the amount of degradation generated over a time interval $(t-\Delta,t]$, $g_{\Delta}(\bm{s},t)$, in the convolution model (\ref{eq:Z}), directly corresponds to the generation term in the physics model (\ref{eqn:transport}) through the following relationship:
$
g_{\Delta}(\bm{s},t) = Q(\bm{s},t)\Delta.
$
And $u(\bm{s},t)$ introduced in (\ref{eq:u}) becomes exactly the degradation rate.

Based on the results above, we now consider a recently proposed reaction-diffusion model that describes the gradual degradation (decomposition) of polymer microspheres composed of poly(D,L-lactic-co-glycolic acid) (PLGA) (Versypt et al., 2015). The PLGA microspheres are used for pharmaceutical drug delivery over extended periods of time. Compared to conventional drug dosage forms, the PLGA microspheres are biodegradable polymeric devices for controlled-release drug delivery, which provide enhanced control of drug concentrations and biodistribution, reduce side effects, and improve patient compliance. The reaction-diffusion model for the degradation of polymer microspheres within a radially symmetric sphere is given by
\begin{equation} \label{eqn:transportnew}
\frac{\partial \varphi(\bm{r},t)}{\partial t} = Q(\bm{r},t) + K \nabla^2 \varphi(\bm{r},t)
\end{equation}
where $\varphi(\bm{r},t)=\bm{r}c(\bm{r},t)$ with $0 \leq \bm{r} \leq 1$ and $c(\bm{s},t)$ respectively being the normalized radial position and concentration, $K$ is the normalized diffusion coefficient, and $Q(\bm{r},t)$ is the net reaction rate of generation of species per volume (interested readers may refer to Versypt et al., (2015) for more details). Note that, (\ref{eqn:transportnew}) is a typical degradation model with a generation and a diffusion term.

Consider a pure diffusion process with $Q(\bm{r},t)=0$, (\ref{eqn:simple_transport}) becomes
\begin{equation}\label{eqn:simple_transportnew}
\frac{\partial \varphi(\bm{r},t)}{\partial t} =  K \nabla^2 \varphi(\bm{r},t).
\end{equation}
Applying the Fourier transform, equation (\ref{eqn:simple_transportnew}) becomes
\begin{equation} \label{eqn:Fourier_transportnew}
\frac{d \tilde{\varphi}_\eta(t)}{d t} = - \eta^2 K \tilde{\varphi}_\eta(t),
\end{equation}
in which $\eta$ is the wave number and $\tilde{\varphi}_\eta$ is the Fourier coefficient of $\varphi$. And from (\ref{eqn:conv_transport}), we have
\begin{equation} \label{eqn:conv_transportnew}
\varphi(\bm{s},t) = \int_{\mathbb{R}^2} \Omega_\Delta(\bm{x}) \varphi(\bm{r}-\bm{x},t-\Delta) d\bm{x},
\end{equation}
in which the kernel $\Omega_{\Delta}(\bm{x})$ is exactly a Gaussian kernel:
\begin{equation}
\Omega_\Delta(\bm{x}) = \frac{1}{2\pi|\bm{\Sigma}_{\Delta}|^{1/2}}\exp\left\{-\frac{\bm{x}^{\top} \bm{\Sigma}_{\Delta}^{-1}\bm{x}}{2}\right\},
\end{equation}
with the covariance matrix given by
\begin{equation} \label{eqn:cov_matnew}
\bm{\Sigma}_\Delta =
\begin{pmatrix}
2 K \Delta & 0\\
0 & 2 K \Delta
\end{pmatrix}.
\end{equation}
Based on (\ref{eqn:conv_transportnew}), the Gaussian convolution kernel can be chosen for the statistical model as shown in (\ref{eq:Z1}) and (\ref{eq:Z2}). Equation (\ref{eqn:cov_matnew}) also suggests that $\rho_1 = \rho_2 = 2K(\bm{r},t)$.

Finally, it follows from (\ref{eqn:conv_approx}) that
\begin{equation} \label{eqn:conv_approxnew}
\Delta \left( \frac{\partial \varphi(\bm{s},t)}{\partial t} - [g_{\Delta}(\bm{s},t) / \Delta] \right) =  O(\Delta^2).
\end{equation}
Hence, as $\Delta \rightarrow 0$, $g_{\Delta}(\bm{s},t)$ in the statistical degradation model is exactly the reaction term in the physical model (\ref{eqn:transportnew}) proposed by Versypt et al.,(2015).

\clearpage
\large{\textbf{References}}

Bae, S. J. and Kvam, P. H. (2004), A Nonlinear Random-Coefficients Model for Degradation Testing. \textit{Technometrics}, \textbf{46}, 460--469.

Banerjee, S., Carlin, B.P. and Gelfand, A.E. (2004), Hierarchical Modeling and Analysis for Spatial Data, Chapman \& Hall/CRC, New York.

Bhattaacharyya, G. K. and Fries, A. (1982), Fatigue Failure Models-Birnbaum-Saunders vs. Inverse Gaussian. \textit{IEEE Transactions on Reliability}, \textbf{31}, 439--440.
  
Bian, L. K., Gebraeel, N. and Kharoufeh, J. P. (2015), Degradation Modeling for Real-Time Estimation of Residual Lifetimes in Dynamic Environments. \textit{IIE Transaction}, \textbf{47}, 471¨C¨C486.

Brouste, A., Istas, J. and Lambert-Lacroix, S. (2007), On Fractional Gaussian Random Fields Simulation. \textit{Journal of
  Statistical Software}, \textbf{23}.

Brown, P.E., Karesen, K.F., Roberts, G.O. and Tonellato, S. (2000), Blur-Generated Non-Separable Space-Time Models. \textit{Journal of the Royal Statistical Society, Ser. B}, \textbf{62}, 847--860.

Calder, C. (2007), Dynamic Factor Process Convolution Models for Multivariate Space-Time Data with Application to Air Quality Assesment. \textit{Environmental and Ecological Statistics}, \textbf{14}, 229--247.

Carroll, R., Chen, E., Li, T., Newton, H., Schmiediche, H. and Wang, N. (1997),
Ozone Exposure and Population Density in Harris County, Texas. \textit{Journal of the American Statistical Association}, \textbf{92},
  392--404.

Chen, N. and Tsui, K. L. (2012), Condition Monitoring and Remaining Useful Life Prediction Using Degradation Signals: Revisited. \textit{IIE Transaction}, \textbf{45}, 939--952.

Cressie, N. and Hawkins, D. (1980), Robust Estimation of the Variogram. \textit{Journal of the International Association for Mathematical Geology}, \textbf{12}, 115--125.

Cressie, N. and Huang, H. (1999), Classes of Nonseparable Spatiotemporal Stationary Covariance Functions. \textit{Journal of the American Statistical Association}, \textbf{94}, 1330--1340.

Doganaksoy, N. and Hall, D. (2013), Gaining Physical Insights from Degradation Data. \textit{Journal of Quality Technology}, \textbf{45}, 188--199.

Doksum, K. A. and Hoyland, A., (1992), Models for Variable-Stress Accelerated Life Testing Experiments Based on Wiener Processes and the Inverse Gaussian Distribution. \textit{Technometrics}, \textbf{34}, 74--82.

Foyer, M. and Zou, D., (2006), A Global Approach for Solving Evolutive Heat Transfer for Image Denoising and Inpainting. \textit{IEEE Transactions on Image Processing}, \textbf{15}, 2558--2573.

Fuentes, M., Chen, L., Davis, J. and Lackmann, G. (2005), A New Class of Nonseparable and Nonstationary Covariance Models for Wind Fields. \textit{Environmetrics}, \textbf{16}, 449--464.

Ghosh, S., Bhave, P., Davis, J. and Lee, H. (2010), Spatio-Temporal Analysis of Total Nitrate Concentrations Using Dynamic Statistical Models. \textit{Journal of the American Statistical Association}, \textbf{105}, 538--551.

Gneiting, T. (2002), Nonseparable Stationary Covariance Functions for Space-Time Data. \textit{Journal of the American Statistical Association}, \textbf{97}, 590--600.

Higdon, D. (2002), Space and Space-Time Modeling Using Process Convolutions. in \textit{Quantitative Methods for Current Environmental Issues}, eds. Anderson, C., Barnett, V., Chatwind, P., and El-Shaarawi, A., Springer Verlag, 37--56.

Higdon, D. (2007), A Process-Convolution Approach to Modeling Temperatures in the North Atlantic Ocean. \textit{Environmental and Ecological Statistics}, \textbf{5}, 173--190.

Hong, Y., Duan, Y., Meeker, W. Q., Stanley, D. L. and Gu, X. (2015), Statistical Methods for Degradation Data with Dynamic Covariates Information and an Application to Outdoor Weathering Data. \textit{Technometrics}, \textbf{57}, 180--193.

Lawless, J. F. and Crowder, M. J. (2004), Covariates and Random Effects in a Gamma Process Model with Application to Degradation and Failure. \textit{Lifetime Data Analysis}, \textbf{10}, 213--227.

Liao, H. T. and Tian, Z. G. (2012),
  A Framework for Predicting the Remaining Useful Life of a Single Unit under Time-Varying Operating Conditions. \textit{IIE Transactions}, \textbf{45}, 964--980.

Lang, A. and Potthoff, J. (2011),
  Fast simulation of Gaussian random fields. \textit{Monte Carlo Methods and Applications}, \textbf{17}, 195--214.

Li, M. and Meeker, W. (2013),
  Application of Bayesian Methods in Reliability Data Analysis. \textit{Journal of Quality Technology}, \textbf{46}, 1--23.

Liu, X. and Tang, L. C. (2013),
  A Bayesian Optimal Design for Accelerated Degradation Tests. \textit{Quality and Reliability Engineering International}, \textbf{26}, 863--875.

Liu, X., Gopal, V. and Kalagnanam, J. (2018),
A Spatio-Temporal Modeling Framework for Weather Radar Image Data in Tropical Southeast Asia. \textit{The Annals of Applied Statistics}, To Appear.

Lu, C. J., Meeker, W. Q. and Escobar, L. A. (1996),
  A Comparison of Degradation and Failure-Time Methods for Estimating a Time-to-Failure Distribution. \textit{Statistica Sinica}, \textbf{6}, 531--546.

Meeker, W. Q. and Escobar, L. A. (1998), \textit{Statistical Methods for Reliability Data}, John Wiley \& Sons, New York.

Meeker, W. Q., Escobar, L. A. and Lu, C. J. (1988),
  Accelerated Degradation Tests: Modeling and Analysis. \textit{Technometrics}, \textbf{40}, 89--99.

Nikulin, M. S., Limnios, N., Balakrishnan, N., Kahle, W. and Huber-Carol, C. (2010), \textit{Advances in Degradation Modeling: Applications to Reliability, Survival Analysis, and Finance}, Springer Science \& Business Media, New York.

Peng, C. Y. and Tseng, S. T. (2010),
  Progressive-Stress Accelerated Degradation Test for Highly-Reliable Products. \textit{IEEE Transactions on Reliability}, \textbf{59}, 30--37.

Reich, B., Eidsvik, J., Guindani, M., Nail, A. and Schmidt, A. (2011), A Class of Covariate-Dependent Spatiotemporal Covariance Functions for the Analysis of Daily Ozone Concentration. \textit{Annals of Applied Statistics}, \textbf{5}, 2425--2447.

Schabenberger, O. and Gotway, C. (2005), \textit{Statistical Methods for Spatial Data Analysis}, Chapman \& Hall/CRC, Boca Raton.

Sigrist, F., Kunsch, H. R. and Stahel, W. A. (2015). Stochastic Partial Differential Equation based Modelling fo Large Space-Time Data Sets. \textit{Journal of the Royal Statistial Society, Ser. B}, \textbf{77}, 3--33.

Sigrist, F., Kunsch, H. and Stahel, W. (2015),
  spate: An R Package for Spatio-Temporal Modeling with a Stochastic Advection-Diffusion Process. \textit{Journal of Statistical Software}, \textbf{63}.

Singpurwalla, N. (1995),
  Survival in Dynamic Environments. \textit{Statistical Science}, \textbf{10}, 86--103.

Stroud, J., Muller, P. and Sanso, B. (2001), Dynamic Models for Spatiotemporal Data. \textit{Journal of the Royal Statistical Society, Ser.B}, \textbf{63}, 673--689.

Tseng, S. T. and Peng, C. Y. (2004), Optimal Burn-In Policy by Using an Integrated Wiener Process. \textit{IIE Transactions}, \textbf{36}, 1161--1170.

Versypt, A. N. F., Arendt, P. D., Pack, D. W. and Braatz, R. D. (2015), Derivation of an Analytical Solution to a Reaction-Diffusion Model for Autocatalytic Degradation and Erosion in Polymer Microspheres. \textit{PLoS
ONE}, \textbf{10}, e0135506, doi:10.1371.

Wikle, C. K. and Cressie, N. (1999), A Dimension-Reduced Approach to Space-Time Kalman Filtering. \textit{Biometrika}, \textbf{86},
  815--829.

Ye, Z. S. and Xie, M. (2014), Stochastic Modelling and Analysis of Degradation for Highly Reliable Products. \textit{Applied Stochastic Models in Business and Industry}, \textbf{31},
  16--32.

Ye, Z. S. and Chen, N. (2014),
  The Inverse Gaussian Process as a Degradation Model. \textit{Technometrics}, \textbf{56}, 302--311.

Zhou, R. S., Serban, N. and Gebraeel, N. (2014),
  Degradation-Based Residual Life Prediction under Different Environments. \textit{The Annals of Applied Statistics}, \textbf{8}, 1671--1689.

\end{document}